\documentclass[%
 reprint,
 nofootinbib,
 amsmath,amssymb,
 aps,
]{revtex4-2}
\usepackage{mathptmx}
\usepackage[english]{babel}
\usepackage{amsmath,amssymb}
\usepackage{upgreek}
\usepackage{xcolor}
\definecolor{darkred}{rgb}{0.5,0,0}
\definecolor{darkblue}{rgb}{0,0,0.5}
\definecolor{firebrick}{rgb}{0.75,0.125,0.125}
\definecolor{darkgreen}{rgb}{0,0.5,0}
\definecolor{deepblue}{rgb}{0,0,0.5}
\definecolor{deepred}{rgb}{0.6,0,0}
\definecolor{deepgreen}{rgb}{0,0.5,0}
\usepackage[colorlinks=true,linkcolor=firebrick,citecolor=darkgreen,urlcolor=darkblue]{hyperref}
\usepackage[capitalise]{cleveref}
\usepackage{xspace}
\usepackage{tabularx}
\usepackage{multirow}
\usepackage{graphicx}\graphicspath{{figs/}}
\usepackage{textgreek}\renewcommand{\upmu}{\text{\textmu}}

\def\Offline{\mbox{$\overline{\textrm{Off}}$\hspace{.05em}\protect\raisebox{.4ex}{$\protect\underline{\textrm{line}}$}}\xspace}
\def\dd{\mathrm{d}}
\def\e{\mathrm{e}}
\def\DX{\Delta X}
\def\Xmax{X_\text{max}}
\def\Rmu{R_\upmu}
\def\Xvg{X_\text{vg}}
\def\gcm{\text{g/cm$^2$}}
\def\hs{h_\text{s}}
\def\hproj{h_\text{proj}}
\def\orcid#1{\href{https://orcid.org/#1}{\includegraphics[height=1.55ex]{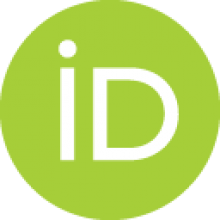}}}
\def\emcomp{{\text{e}\upgamma}}
\def\mucomp{{\upmu}}
\def\emmucomp{{\text{e}\upgamma(\upmu)}}
\def\emhdcomp{{\text{e}\upgamma(\uppi)}}
\def\un2{\texttt{Universality\,II}}

\begin{document}

\title{A Model of the Response of Surface Detectors to Extensive Air Showers Based on Shower Universality}

\author{Maximilian Stadelmaier\,\orcid{0000-0002-7943-6012}}

\affiliation{Institute for Astroparticle Physics, Karlsruhe Institute of Technology \includegraphics[height=1.55ex]{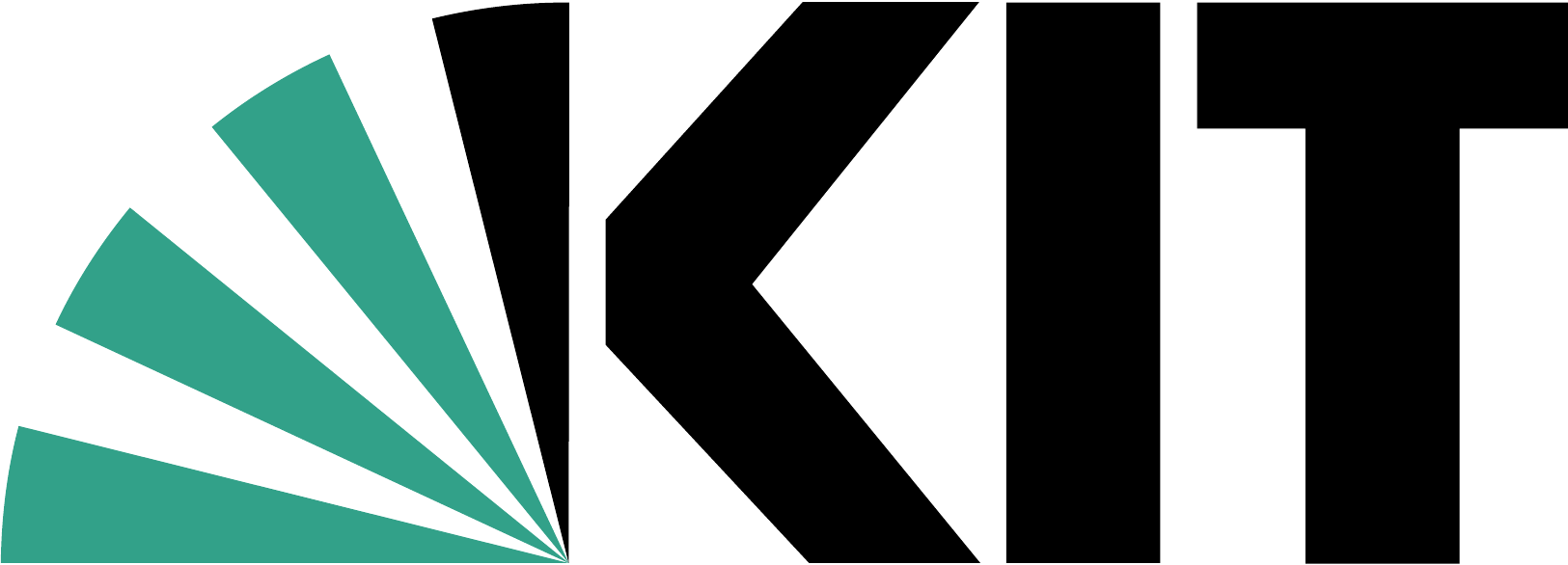}, Karlsruhe, Germany}
\affiliation{Institute of Physics of the Czech Academy of Sciences \raisebox{-0.2\height}{\includegraphics[height=2.2ex]{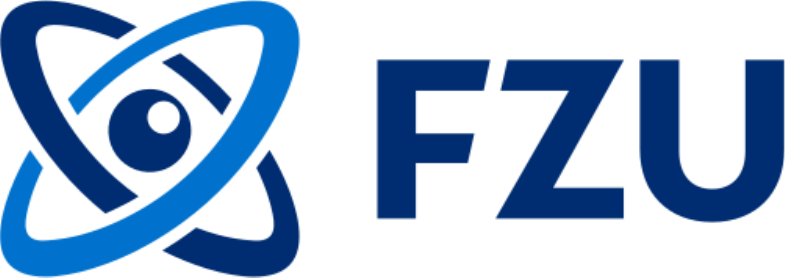}}, Prague, Czech Republic}
\affiliation{Istituto Nazionale di Fisica Nucleare, \raisebox{-0.2\height}{\includegraphics[height=2.2ex]{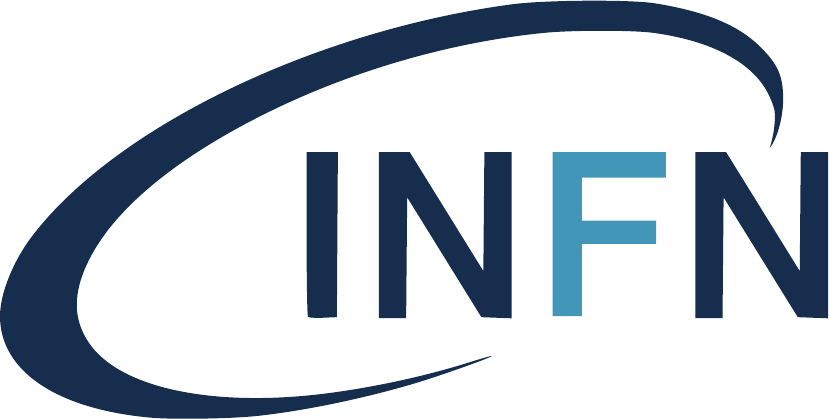}} Milano, Italy}

\author{Ralph Engel\,\orcid{0000-0003-2924-8889}}
\affiliation{Institute for Astroparticle Physics, Karlsruhe Institute of Technology \includegraphics[height=1.55ex]{kit_logo-no_text}, Karlsruhe, Germany}

\author{Markus Roth\,\orcid{0000-0003-1281-4477}}
\affiliation{Institute for Astroparticle Physics, Karlsruhe Institute of Technology \includegraphics[height=1.55ex]{kit_logo-no_text}, Karlsruhe, Germany}

\author{David Schmidt\,\orcid{0000-0001-6963-1191}}
\affiliation{Institute for Astroparticle Physics, Karlsruhe Institute of Technology \includegraphics[height=1.55ex]{kit_logo-no_text}, Karlsruhe, Germany}

\author{Darko Veberi\v{c}\,\orcid{0000-0003-2683-1526}}
\affiliation{Institute for Astroparticle Physics, Karlsruhe Institute of Technology \includegraphics[height=1.55ex]{kit_logo-no_text}, Karlsruhe, Germany}

\begin{abstract}
We present a full model of surface-detector responses to extensive air showers.
The model is motivated by the principles of air-shower universality and can be applied to different types of surface detectors. 
Here we describe a parametrization for both water-Cerenkov detectors and scintillator surface detectors, as for instance employed by the upgraded detector array of the Pierre Auger Observatory.
Using surface detector data, the model can be used to reconstruct with reasonable precision shower observables such as the depth of the shower maximum $\Xmax$ and the number of muons $\Rmu$.
\end{abstract}

\keywords{Ultra high-energetic cosmic rays --- extensive air showers --- air-shower universality --- Pierre Auger observatory}

\maketitle

\section{Introduction}
\label{sec:intro}

Ultra-high-energy cosmic rays (UHECRs) are the highest energy particles known to humankind; they challenge the hypothesized limits for the acceleration of particles, while a single source of UHECRs has not yet been identified~\cite{AlvesBatista:2019tlv}.
Although experimental evidence implies that the spectrum of cosmic rays is strongly dominated by ionized nuclei for energies greater than $10^{18}$\,eV~\cite{PierreAuger:2019ens,PierreAuger:2020kuy}, the exact chemical composition of the UHECRs is still unknown.

Due to the low flux at the highest energies, the UHECRs are not observed directly, but indirectly by the extensive air showers (EASs) they induce in the Earth's atmosphere.
These are cascades of secondary particles created when a UHECR interacts with an air nucleus at the top of the atmosphere.
At the highest energies, the footprint of an EAS at the ground reaches several kilometers in diameter and can therefore be detected by a sparse array of surface detector (SD) stations.

To estimate the chemical composition (nuclear masses) of UHECRs, a precise understanding of the air shower phenomenon is crucial.
Especially the depth $\Xmax$ of the maximum of air-shower development and the relative number of muons $\Rmu$ in the shower are strong indicators of the primary mass of the cosmic ray.
Studies on both $\Xmax$ and $\Rmu$ individually have been performed successfully~\cite{pierreaugercollaboration2014depth,PierreAuger:2014ucz,PierreAuger:2021qsd}.

$\Xmax$ is an observable that is obtained from direct observation of the longitudinal profile of fluorescence light produced by a shower, which is, however, only possible during clear moonless nights.
The shower observed in fluorescence light appears brightest at the point where its development reaches the maximum $\Xmax$.
Using only SD data, $\Xmax$ can be estimated indirectly from the curvature and thickness of the shower front reaching the ground~\cite{PierreAuger:2017tlx}.

The relative\footnote{In this article, we will use the average muon content of showers induced by proton primary cosmic rays simulated with the \textsc{Epos-LHC}~\cite{Pierog:2013ria} model of hadronic interactions as a reference. A quantitative definition will be given later in the text.} muon number $\Rmu$ can be estimated using the signal of water-Cherenkov detectors (WCDs) such as those employed by the Pierre Auger Observatory (Auger, in short), if the primary energy of the cosmic ray can be estimated independently.
This can be achieved, for example, with a fluorescence detector or with the help of scintillator surface detectors (SSD), such as those deployed in the AugerPrime upgrade~\cite{PierreAuger:2016qzd}.

In this work, we present a model of the expected particle densities in air showers for given values of $\Xmax$ and $\Rmu$ at a given primary energy.
The model is based on the concept of ``air-shower universality''~\cite{Hillas:1982vn}, or in short \emph{Universality} (for an overview see Ref.~\cite{Lipari:2008td}). 
It describes the dependence of the expected signal in the SD as a function of the atmospheric distance to the shower maximum, which is otherwise considered only on average using the constant-intensity-cut method; it describes the lateral evolution of the shower particle densities; and it describes the temporal distribution of shower particles at the ground.
The parameters of the model are determined using Monte Carlo simulations of the signal deposited in an Auger-like SD array using WCDs and SSDs.
Using the observables $\Xmax$ and $\Rmu$ as free parameters in a fit, the model can be used to describe the detector data and is thus able to estimate the mass of the primary particle.

This work builds on previous implementations of the universality model~\cite{aveICRC,AVE201723,Ave201746,phdmaurel,phdschulz,Bridgeman:2018nsw,Hulsman:2019zmg} in the \Offline software framework~\cite{Argiro:2007qg} of the Pierre Auger Observatory.
In contrast, the new implementation introduces physically-motivated shower profile functions, as well as full detector simulations for the final parametrizations.
The model developed in this work we will refer to as \un2.

\section{Extensive air showers}

An EAS is a cascade of subatomic particles created upon impact of an UHECR with the Earth's atmosphere.
The particles in an air shower multiply through collisions, decay, and radiation until the energy budget of the primary cosmic ray is converted entirely.

\begin{figure}
    \centering
    \includegraphics[width=0.9\columnwidth]{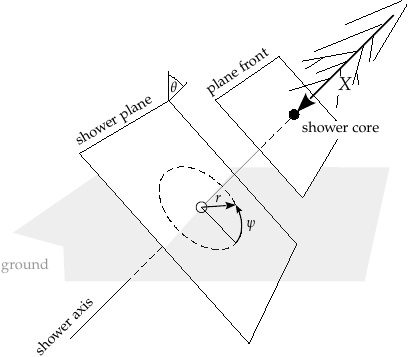}
    \caption{Illustration of the event geometry.
    The shower core at the depth $X$ is depicted as full black marker.
    The plane front of the shower contains the shower core and is perpendicular to the shower axis.
    The impact point of the shower core at the ground is given as hollow circular marker.
    The shower plane contains the shower core at the ground and is perpendicular to the shower axis. The zenith angle $\theta$ describes the inclination of the shower axis.
    Positions of points on the ground are expressed with their shower-plane coordinates $r$ and $\psi$.}
    \label{fig:geom}
\end{figure}

Most particles are created in the core of the EAS, which develops along the extended trajectory of the primary UHECR and which defines the shower \emph{axis} as a line of reference; the plane that lies perpendicularly to the shower axis and that contains the current location of the shower core is defined as the \emph{plane front} of the shower.
The propagation of the plane front and the core is usually described as a function of slant atmospheric depth $X$ (in units of $\gcm$).
\emph{Shower plane} is defined as the plane front that contains the impact point of the shower core on the ground.
The positions of the detectors on the ground may be given in terms of their polar coordinates in the shower plane, defined by the distance $r$ to the shower axis and the angle $\psi$ between the station and the direction in which the shower axis is tilted.
The inclination of the shower axis with respect to the zenith is described by the zenith angle $\theta$.
The geometry of the event is illustrated in \cref{fig:geom}.

Even though the particle content of an EAS is heavily governed by the hadronization occurring in the first interactions, most of the particles in the cascade are created by electromagnetic processes and thus most of the energy is deposited in the atmosphere by electrons, positrons, and photons. 
Muons, which are mostly created from the decay of charged pions, play a significant role in the detection of UHECRs, too.
Firstly, they can be easily detected by surface detectors even in inclined events, because of their long lifetime and smaller energy losses. 
Secondly, they provide a measure for the amount of charged pions created in the shower, and thus of hadronization that took place in the first interactions~\cite{Cazon:2018gww,PierreAuger:2021qsd}.
Since the hadronization of the first interactions is enhanced by the number of nucleons taking part in the first interaction, the amount of muons in an EAS is an indicator for the nuclear mass of the primary particle.
The number of particles created by hadronic and electromagnetic processes is schematically described in the Heitler-Matthews model of hadronic showers~\cite{Matthews:2005sd}.
In this model, a shower is initiated by a primary proton interacting hadronically with the atmosphere.
In every hadronic interaction, charged and neutral pions\footnote{The creation and interaction of heavier mesons are neglected.} are created in a 2:1 ratio.
In this model, charged pions will continue to interact hadronically, while neutral pions promptly decay into photons, fueling the hadronic and electromagnetic cascades.
The energy is always distributed equally among all particles created in an interaction.
Once the charged pions fall below a critical energy of $\epsilon_\text{c}^\uppi \simeq 20$\,GeV, they stop fueling the cascade and decay into muons.
The total number of muons $N_\upmu$ created in a shower induced by a proton of energy $E_0$ thus scales as
\begin{equation}
    N_\upmu = \left(\frac{E_0}{\epsilon_\text{c}^\uppi}\right)^\beta,
    \label{eq:nmu_hm}
\end{equation}
where the relative muon growth rate is $\beta\lesssim1$.

According to the Heitler model~\cite{Heitler:1936jqw} of electromagnetic cascades, the electromagnetic particles multiply without any hadronic interaction until a critical energy of $\epsilon^\emcomp_\text{c}~\simeq~87$\,MeV is reached.
At this point, the size of the shower, that is, the number of particles present, is at its maximum.
The slant atmospheric depth of the position of the shower maximum is referred to as $\Xmax$.

Again according to the Heitler-Matthews model, in showers initiated by primary nuclei with nuclear mass $A$, all nucleons initiate individual and independent cascades, each of them with a $1/A$ fraction of the total energy of the primary nucleus, while ignoring shower-to-shower fluctuations.
Thus, the critical energy $\epsilon_\text{c}^\uppi$ is reached after fewer interactions for primary nuclei with $A > 1$ than for protons ($A=1$).
This implies that on average the depth $\Xmax$ is smaller for heavier primary particles.
Furthermore, because the relative rate of muon growth $\beta$ in \cref{eq:nmu_hm} is less than one, on average a factor of $A^{1-\beta}$ more muons are produced in showers induced by primary nuclei with mass $A$ than in proton showers.

In the context of the mass composition of UHECRs, the depth of the shower maximum $\Xmax$ and the number of muons $N_\upmu$ created in a shower are thus important mass-sensitive observables.

\section{A Model from Air-Shower Universality}
\label{sec:univ}

\begin{figure}
    \centering
    \def\w{0.9}
    \includegraphics[width=\w\columnwidth]{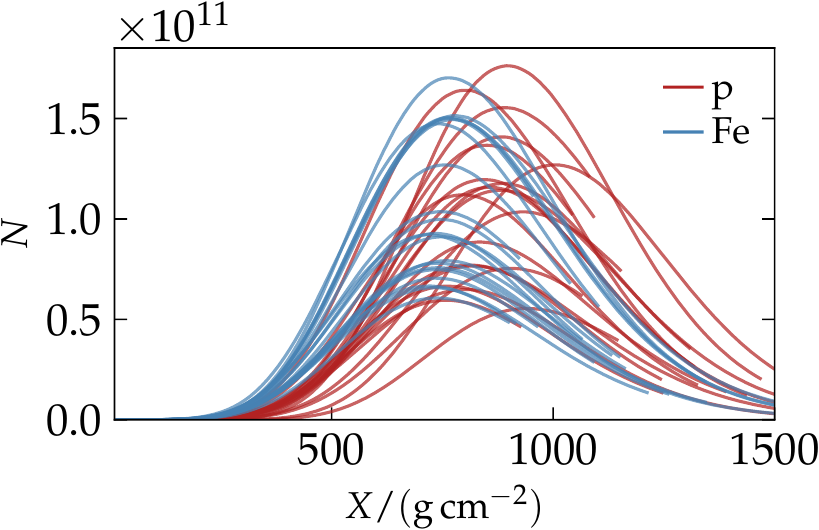}
    \\[3mm]
    \includegraphics[width=\w\columnwidth]{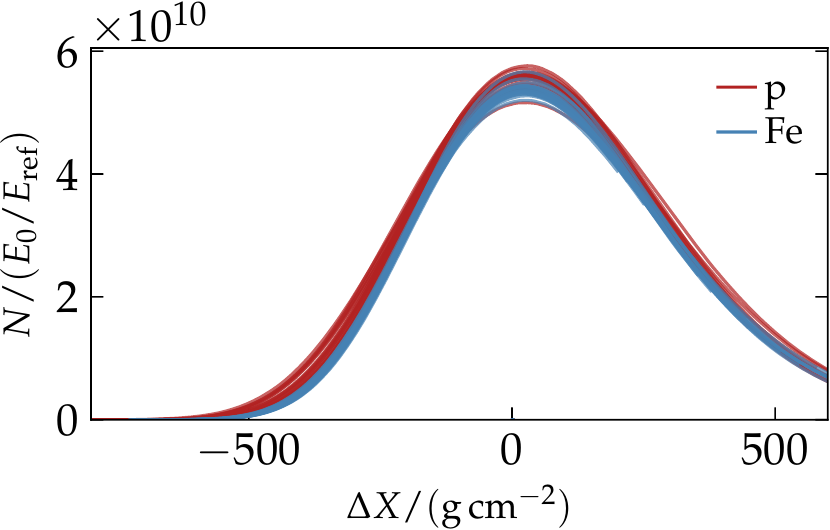}
    \caption{Longitudinal number profiles $N(X)$ of simulated air showers from proton and iron primary particles, p and Fe, as a function of depth $X$.
    The showers are simulated with primary energies between $10^{19.5}$ and $10^{20}$\,eV and with different zenith angles.
    Using the primary energy $E_0$ and reference energy $E_\text{ref}=10^{19}$\,eV, the $N(X)$ profiles in the second panel are scaled by $E_0/E_\text{ref}$ in magnitude and shifted individually, $\DX=X-\Xmax$, to place the maximum at zero and align the various showers.}
    \label{fig:un_lon}
\end{figure}

Air-shower universality, as first observed in simulations by Hillas~\cite{Hillas:1982vn}, states that the electromagnetic particles in individual air showers initiated by primary particles with a large energy $E_0$ show the same energy spectrum, as well as radial and angular distributions, and that these distributions and spectra are well described by idealized models such as those given in Refs.~\cite{RossiGreisenCRTheory,Nishimura:1950lateral,lafebre2009universality}.
Especially around the maximum of the shower, where the electromagnetic particles of a shower on average carry the same energy $\epsilon_\text{c}^\emcomp$, the longitudinal and lateral distributions of the particles (profiles), the energy spectra, and the angular distributions of particles are in a good approximation the same for all showers initiated with the same primary energy $E_0$.
This is true independently of the absolute value of $\Xmax$ and of the type of primary particle.
Simulated longitudinal profiles of the showers, which describe the number of particles $N$ present at depth $X$, are depicted in \cref{fig:un_lon}.
Profiles become universal when introducing a shower-depth parameter $\DX = X-\Xmax$, which describes the depth relative to the shower maximum, and when the number of particles is simultaneously scaled with the primary energy; $10^{19}$\,eV is chosen as the reference energy $E_\text{ref}$, to which the profiles shown in \cref{fig:un_lon} are scaled.

The longitudinal profile of air showers and the spectra of the corresponding electromagnetic particles are strongly related to each other.
It was already demonstrated analytically in the 1930s~\cite{RossiGreisenCRTheory} and later using intense simulations~\cite{lafebre2009universality} that the spectra of electromagnetic particles in air showers can be well described as a function of the shower-age parameter $s$.
The parameter $s$ can be expressed as a function of depth $X$ and depth of the shower maximum $\Xmax$ approximately as $s \simeq 3 X / (X + 2 \Xmax)$~\cite{1956progress}.
The relative rate of change of the longitudinal profiles\footnote{An example function for $g$ is given in \cref{eq:mGH}.}, $g$, of electromagnetic particles, $\lambda_1=(\partial g/\partial X)/g$, is well described as a function\footnote{Possible functions for $\lambda_1$ are given in~\cref{app:rel}.} of $s$~\cite{RossiGreisenCRTheory,Lipari:2008td}, which unambiguously relates the longitudinal profile $g$ to the shower age.

Furthermore, for depths $X$ around $\Xmax$, the NKG function~\cite{Greisen:1960}, which has been proven to successfully describe the lateral distribution of shower particles~\cite{KASCADE2001}, is approximately constant in $s$.
Therefore, around the maximum, the dependence of the particle distributions on energy, depth, and radius factorizes.
Furthermore, the number of electromagnetic particles for any primary energy $E_0$ is well described by a power law, $(E_0/E_\text{ref})^\gamma$ with $\gamma\lesssim1$~\cite{Matthews:2005sd}.
Thus, in general, the density of electromagnetic particles in the shower plane at the depth $\DX$ near the shower maximum and at the distance $r$ to the shower axis can be factorized to a good approximation by a universal function of the form~\cite{1956progress}
\begin{equation}
    \varrho \simeq (E_0/E_\text{ref})^\gamma \; f(r) \; g(\DX),
    \label{eq:masterequation}
\end{equation}
with a longitudinal profile function $g(\DX)$ and an NKG-like function $f(r)$;
the zenith-angle dependence of the number of particles reaching the SD is effectively described by $g(\DX)$ with the distance $\DX$ to the shower maximum, cf.\ \cref{eq:un_dx_def}.
This concept has been extended to different particle components in air showers~\cite{AVE201723}, so that not only electromagnetic, but also hadronic showers can be well described by a model based on Universality.
For this purpose, the particles in air showers are subdivided into four components, for which the longitudinal and lateral profiles are parametrized individually using a function of the form given in \cref{eq:masterequation}.
The number of particles in each component -- except for the electromagnetic -- scales approximately linearly with the relative amount of muons $\Rmu$ in the shower.
In contrast, the size of the electromagnetic component is to a large extent independent of $\Rmu$.
This is discussed in \cref{app:emscale}.
The total shower content is then comprised by the sum of the four components.
Since $\Rmu$ is in general an observable that depends on the primary particle, it is expected that three of the four particle components will be systematically improved in showers initiated by heavier primary particles. 

The four-component model, as well as the longitudinal and lateral profiles used to parametrize the individual particle densities, will be discussed in the following.

\subsection{The four-component shower model}
\label{sec:four_components}

We consider hadronic air showers to be constituted by the four particle components introduced in~\cite{AVE201723}.
The resulting shower is disentangled into each of the four components with different longitudinal and lateral profiles.
The electromagnetic component, $\emcomp$, contains electrons and positrons, as well as photons, all of them created by the electromagnetic cascade of the air shower.
The muon component, $\mucomp$, contains all muons and antimuons from the shower.
The electromagnetic muon component, $\emmucomp$, contains electromagnetic particles that were created by decaying particles of the muon component in the first or second generation.
The hadronic component, $\emhdcomp$, contains all electromagnetic particles that were created in hadronic decays within two generations as well as all hadrons.
The rest of the particles, such as neutrinos, which do not deposit a significant amount of signal in a surface detector, are neglected.

The number of particles $N_i$ of the $i$-th component is considered at a fixed depth $\DX_\text{ref}$ in the shower plane.
The amount of particles relative to the respective average in a proton shower $\langle N_i^\text{p}\rangle$ is thus given by
\begin{equation}
     R_i := \frac{N_i}{\langle N_i^\text{p}\rangle},
     \label{eq:Ri}
\end{equation}
which for $i=\upmu$ defines the relative number of muons.
For convenience, the size of all particle components is expressed as a function of $\Rmu$, independent of the energy.
By definition, the proton showers have on average $\Rmu=1$.
In the first order, the size of each component scales linearly w.r.t.\ $\Rmu$, according to the relation~\cite{AVE201723}
\begin{equation}
    R_i - 1 = a_i\,(\Rmu - 1),
\label{eq:compcorrelation}
\end{equation}
with a constant $a_i$ for each particle component.
Since the number of particles in the electromagnetic component scales roughly independently of $R_\upmu$, $a_\emcomp$ is expected to be very close to 0, but slightly smaller than 0 due to energy conservation.
For the $\emmucomp$ and $\emhdcomp$ components values of $a_\emmucomp \simeq a_\emhdcomp \simeq 1$ are expected.
Except for their correlation according to \cref{eq:compcorrelation}, the four components are treated individually and independently.

For each component, the total number of particles as a function of the primary energy $E_0$ is given by a power law.
For reference, the expected number of particles produced by a proton shower with $E_0=10^{19}$\,eV is written as $N_{19,i}$.
The expected number of particles in a shower of primary energy $E_0$ is thus given by
\begin{equation}
    N_i(E_0) = \,N_{19,i} \left(\frac{E_0}{10^{19}\,\text{eV}}\right)^{\gamma_i}.
    \label{eq:Ni}
\end{equation}
For better readability we will in the following drop the component index $i$ from all component-specific parameters.

\subsection{The longitudinal profile}
\label{subsec:long}

\begin{figure*}
    \centering
    \def\h{0.39}
    \includegraphics[height=\h\textwidth]{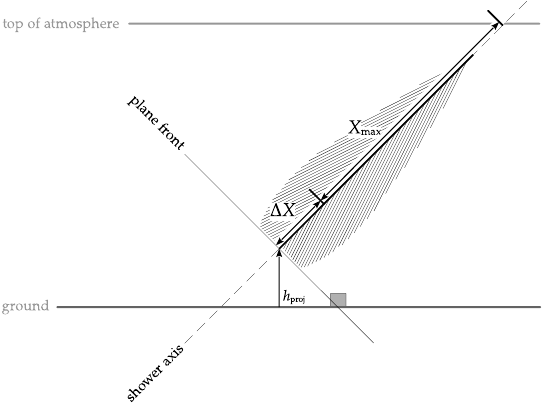}\hfill
    \includegraphics[height=\h\textwidth]{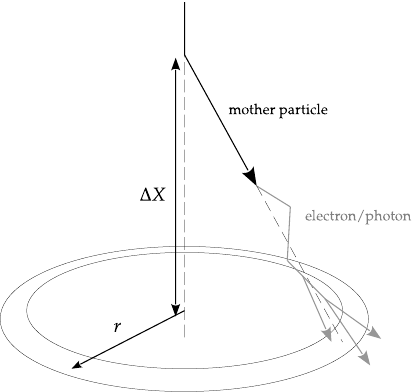}
    \caption{\emph{Left:} Schematic of an event geometry.
    The axis of the shower is shown as a dashed line, and the path of the shower core is shown as a bold line.
    The slant atmospheric depth $\Xmax$ of the shower maximum is measured from the top of the atmosphere along the shower axis.
    The depth parameter $\DX$ of a detector (gray box at the ground) is measured along the shower axis between shower maximum and the perpendicular (plane-front) projection of the detector onto the shower.
    The (physical) height of the shower core at depth $\DX$ is given by $\hproj$.
    \emph{Right:} Path of a mother particle (black arrow), produced close to the shower core and the shower maximum, which decays into electrons and photons (gray arrows).
    The resulting $\emmucomp$ and/or $\emhdcomp$ components are therefore originating from depth $\DX$ and will hit the ground at a distance $r$ from the shower axis (rings); image reproduced from Ref.~\cite{AVE201723}.}
    \label{fig:dx}
\end{figure*}

The Gaisser-Hillas profile function accurately describes the longitudinal development of EAS, as was directly measured in Ref.~\cite{Aab_2019}.
To parametrize the particle density in any point at a distance $r$ and at a depth\footnote{The relation between shower-plane coordinates of a detector and the depth $\DX$ is given in \cref{eq:un_dx_def}.} $\DX$, the modified Gaisser-Hillas profile was introduced in Ref.~\cite{AVE201723}.
A sketch of the geometry is given in \cref{fig:dx} (\emph{left}).
We use the modified Gaisser-Hillas function to describe the development of the particle density in the shower plane front along the shower axis.
At the moment when the plane front intersects with a surface detector, the shower core is at the height $\hproj$ (cf.\ \cref{eq:hproj}) above the ground.
The modified Gaisser-Hillas profile $g(\DX)$ is normalized to a fixed depth $\DX_\text{ref}$ after the maximum of the shower. 
For each of the four components of the shower, the particle density can be parametrized using $g(\DX)$ as
\begin{align}
  \varrho &= \varrho_\text{ref}(r) \, g(\DX),
\\
  g(\DX) &=
    \left[\frac{\DX - \DX_1}{\DX_\text{ref}-\DX_1}\right]^{\frac{\DX_\text{max} - \DX_1}{\lambda}}
    \exp\left[-\tfrac{\DX - \DX_\text{ref}}{\lambda}\right].
\label{eq:mGH}
\end{align}
The reference density $\varrho_\text{ref}$ at the depth $\DX_\text{ref}$ will be discussed in the next section.
For the model described in Section~\ref{sec:param}, $\DX_\text{ref}$ is set to $200\,\gcm$ for all components, which is approximately the depth at which the surface detector is expected for most showers (the median zenith angle is $\theta \simeq 38^\circ$).
Parameters $\lambda$, $\DX_1$, and $\DX_\text{max}$ are individually parameterized for each component using simulations.
As in the original Gaisser-Hillas profile, $\lambda$ is related but not equal to the radiation length of the individual particles\footnote{For the $\emcomp$ component it is expected to be $\lambda \simeq 3 X_0 / 2$, where $X_0$ is the electromagnetic radiation length in air.}.
The parameter $\DX_1$ marks the effective start of the cascade for a component and is fixed to large negative constant values of ${\approx}-500\,\gcm$.
Since Universality is only observed at and after the shower maximum, the values of $\DX_1$ have no physical meaning and are not directly related to the first interaction.
$\DX_\text{max}$ parametrizes the radial retardation of the shower maximum, which could be equivalently described by a distance-dependent shower age~\cite{1956progress,Linsley:1978xw}.
Especially for the components $\mucomp$ and $\emmucomp$, the effective shower maximum is shifted towards $\DX > 0$ and increases with $r$.
Both $\lambda$ and $\DX_\text{max}$ are parametrized as constant or linear functions in $r$.

\subsection{The lateral reference profile}
\label{subsec:lat}

The lateral distribution of particles in air showers is well described by the functions that were derived in Refs.~\cite{Nishimura:1950lateral,Nishimura:1952}, and later approximated in Ref.~\cite{Greisen:1960}.
For a shower with primary energy $E_0$, the reference density $\varrho_\text{ref}$, which describes the shower at a fixed depth $\DX_\text{ref}$ with respect to the maximum of the shower, can be parametrized using the NKG function as $f(r)$ and \cref{eq:Ni}, resulting in
\begin{equation}
  \varrho_\text{ref}(r) = (E_0/E_\text{ref})^\gamma \; f(r)
  \label{eq:rhoref}
\end{equation}
with $E_\text{ref} = 10^{19}$\,eV and
\begin{equation}
  f(r) =  f_0
            \left( \frac{r}{r_\text{G}}\right)^{s-2}
            \left( 1 + \frac{r}{r_\text{G}}\right)^{s-\tfrac{9}{2}},
  \label{eq:NKG}
\end{equation}
where
\begin{equation}
    f_0 = N_{19}\,\frac{\Gamma(4.5 - s)}{2\pi r_\text{G}^2\,\Gamma(s)\,\Gamma(4.5-2s)}
\end{equation}
is used for the normalization.
Since $\varrho_\text{ref}$ is always evaluated at $\DX = \DX_\text{ref}$, the parameter $s$ as well as the reference distance $r_\text{G}$ are fixed for each component.\footnote{Because of the increasing thickness of the slanted atmosphere, this description is not valid for very inclined showers with $\theta \gtrsim 60^\circ$, for which $r_\text{G}$ would have to be adjusted as a function of $\sec\theta$~\cite{1956progress}.}
The dependence of the lateral distribution of particles with respect to $\DX$ is well described by the parametrization of $\lambda$ and $\DX_\text{max}$ as a function of the shower-plane radius $r$ (see Section~\ref{subsec:long}).

Due to changes in the opening angle of the particles relative to the the shower axis and due to the attenuation and effects from a non-homogeneous atmosphere, an additional correction factor $u(\psi)$ has to be applied to \cref{eq:NKG}.
A possible parameterization of $u(\psi)$ is given in \cref{eq:un_c_psi}.

\subsection{The areal density of particles}

Combining the scaling of the size of the individual components according to \cref{eq:Ri} with the longitudinal and lateral profile functions $g$ and $f$, discussed in \cref{subsec:long} and \cref{subsec:lat}, as well as with a correction factor $u$ for azimuthal asymmetry, a model of the particle density for each component according to \cref{eq:masterequation} can be given.
The particle density of each component as a function of $\DX$ and $\Rmu$~\cite{Ave201746} reads as
\begin{equation}
\varrho =
  \left(a\, (\Rmu - 1) + 1\right) \,
  \left(\tfrac{E_0}{E_\text{ref}}\right)^\gamma \,
  g(\DX) \;
  f(r) \;
  u(\psi),
\label{eq:un_rho_tot}
\end{equation}
where, again, indices $i$ have been dropped for legibility.
The total density of the particles is given by
\begin{equation}
    \varrho_\text{tot} = \sum\limits_i \varrho_i,
\end{equation}
with the sum over the four particle components $i$.
To describe the detector responses in an SD array, \cref{eq:un_rho_tot} is parametrized directly in terms of the signal for each component, since the Greisen factorization given in \cref{eq:masterequation} also holds in terms of signals.
The parameterization of the signal components according to \cref{eq:un_rho_tot} is discussed in Section~\ref{sec:param}.

\subsection{The temporal distribution of particles}
\label{subsec:time}

The temporal distributions of particles at the ground depend on the geometry of the event.
Particles are considered to propagate with approximately the speed of light and to be mostly created in the shower core.
If particles, for example, are created in a point-like source above the ground, their arrival times would reflect an expanding sphere; if particles are created at an infinite distance, their arrival times would be in all detectors in coincidence with the arrival of the plane front.
Assuming that particles reaching an SD arrive time-ordered with respect to their creation in the shower core, the temporal distribution of particles (and thus the signal) in each detector can be related to the longitudinal profile of the shower~\cite{PierreAuger:2014zay} (for this reason we express all points in time relative to the time of the plane front $t_\text{pf}$ reaching a detector station).

We assume that the distribution of the production depths of particles along the shower axis is similar to the longitudinal profile of the shower.
This has been proven especially for the muonic component of the shower in Ref.~\cite{Cazon:2012ti}.
Assuming a Gaisser-Hillas-like longitudinal profile, the $40\%$ time quantile $t_{40}$ of the signal deposited in a detector is directly related to $\Xmax$, since approximately $40\%$ of the integrated longitudinal profile is contained between $X=X_1$ and $X=\Xmax$~\cite{stadelmaier:2022}.
Furthermore, assuming that particles are produced in the shower core along the shower axis, the curvature of the shower front implicitly carries information about the depth of the shower maximum.
We use a simplified ansatz, assuming that particles that reach an SD at the shower-plane distance $r$ propagate rectilinearly over a distance of
\begin{equation}
    \ell = \sqrt{\smash[b]{D_{X'}^2 + r^2}}.
\end{equation}
Assuming an isothermal atmosphere, with vertical depth of the ground $X_\text{vg}$ and scale height $\hs$, we try to identify the distance $D_{X'}$, which is the distance the shower plane travels from the point of origin of the particle at a depth $X'$ until the shower plane passes through a detector.
Depth $X'$ can be expressed as
\begin{align}
  X' &\simeq
  \frac{\Xvg}{\cos\theta} \,
  \exp\left(-\frac{D_{X'} \, \cos\theta}{\hs} - \frac{\hproj}{\hs}\right)
\\
  &= (\DX+\Xmax) \, \exp\left(-\frac{D_{X'} \, \cos\theta}{\hs}\right),
\end{align}
and therefore
\begin{equation}
  D_{X'} \simeq
  \frac{\hs}{\cos\theta} \,
  \ln\left(\frac{\DX+\Xmax}{X'}\right).
\label{eq:Dxt40}
\end{equation}
The time quantile $t_{40}$, which is assumed to be related to the shower maximum, i.e.\ $D_{\Xmax}$, can thus be expressed as\footnote{The arrival time of the first particles at a detector station at the distance $r$ in the plane front at the depth $\DX$ is approximately related to the depth $X_1$ of the first interaction by the same relation evaluated at $X=X_1$.}
\begin{equation}
  c \, t_{40} \simeq \sqrt{\smash[b]{D_{\Xmax}^2 + r^2}}- D_{\Xmax}
\label{eq:t40}
\end{equation}
for particles propagating on straight trajectories with the speed of light $c$.
We verified that $t_{40}$ is directly dependent on $\Xmax$ in the way implied by \cref{eq:t40,eq:Dxt40}.

In Ref.~\cite{Ave201746} it has been demonstrated that a shifted log-normal distribution function successfully describes the temporal distribution of the signal in an SD.
Furthermore, the functional forms derived in Ref.~\cite{Cazon:2004zx}, which relate the production of muons according to a longitudinal profile along the shower axis with the times of muons arriving at the ground, can also be numerically approximated by a shifted log-normal distribution as well.
The shift is given by the start time of the signal with respect to the time of the plane front passing the detector.
Therefore, the temporal distribution of $n_\text{tot}$ particles in each component arriving at an SD is modelled using a shifted log-normal distribution function $\mathcal{N}_\ell$, which can be expressed in terms of $t_{40}$, 
\begin{align}
  \frac{\dd}{\dd t} n &=
  n_\text{tot} \, \mathcal{N}_\ell(t-t_0, \sigma^2)
\label{eq:un_log_normal_t40}
\\
  &= \frac{n_\text{tot}}{\sqrt{2\uppi}\,\sigma\,(t-t_0)} \,
     \exp\left(-\frac{1}{2 \sigma^2}\left[\ln\left(\tfrac{t-t_0}{t_{40}-t_0}\right) + k\right]^2\right),
\nonumber
\end{align}
using\footnote{If the median time $t_{50}$ is used, the argument of the inverse error function is 0 and $k=0$.}
\begin{equation}
k =
  \sqrt{2}\sigma\,\operatorname{erf}^{-1}(2{\times}\tfrac{40}{100}-1)
  \approx -0.253\sigma.
\end{equation}
The start time $t_0$ and the time quantile $t_{40}$ are given relatively to the time $t_\text{pf}$ of the plane front passing through the detector, which itself is defined by the reconstructed event geometry.
The shape parameter $\sigma$ and $t_{40}$ are parameterized as a function of $\DX$ using detector simulations.
$\sigma$ is expressed as a function of the mean and standard deviation of the arrival times of the particles, which both can be approximated by a linear function in $\DX$, and $t_{40}$ is given by a function of the form given in \cref{eq:t40}.
Furthermore, for $t \leq t_0$, where \cref{eq:un_log_normal_t40} is not well defined, $\dd n / \dd t = 0$.

\begin{figure*}
    \centering
    \def\w{0.46}
    \includegraphics[width=\w\textwidth]{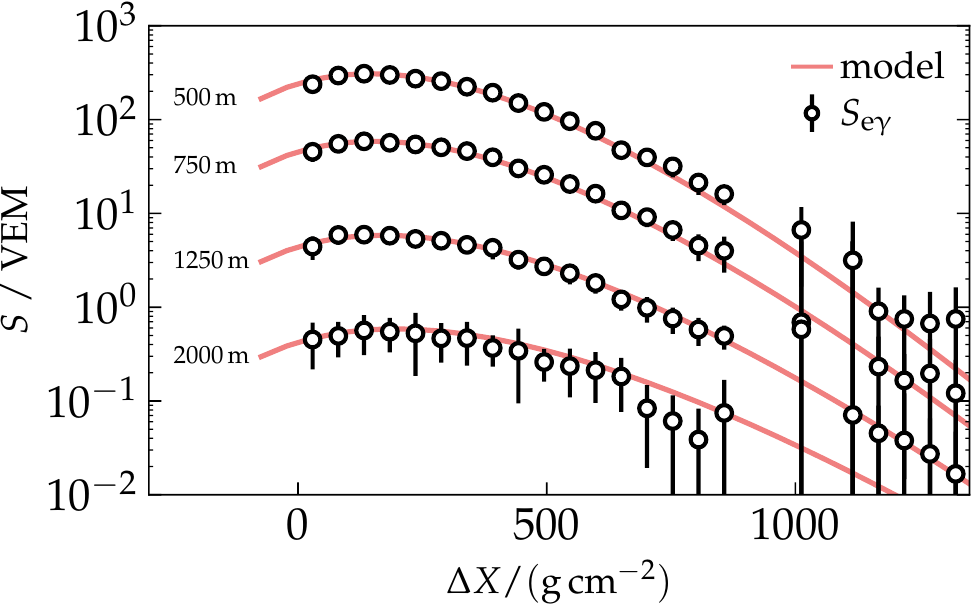}\hfill
    \includegraphics[width=\w\textwidth]{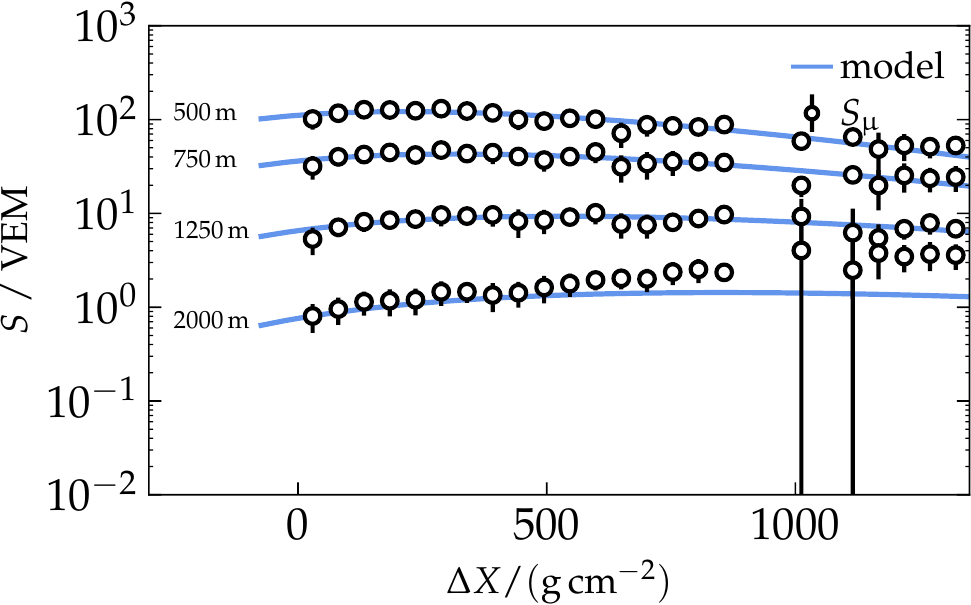}
    \\[3mm]
    \includegraphics[width=\w\textwidth]{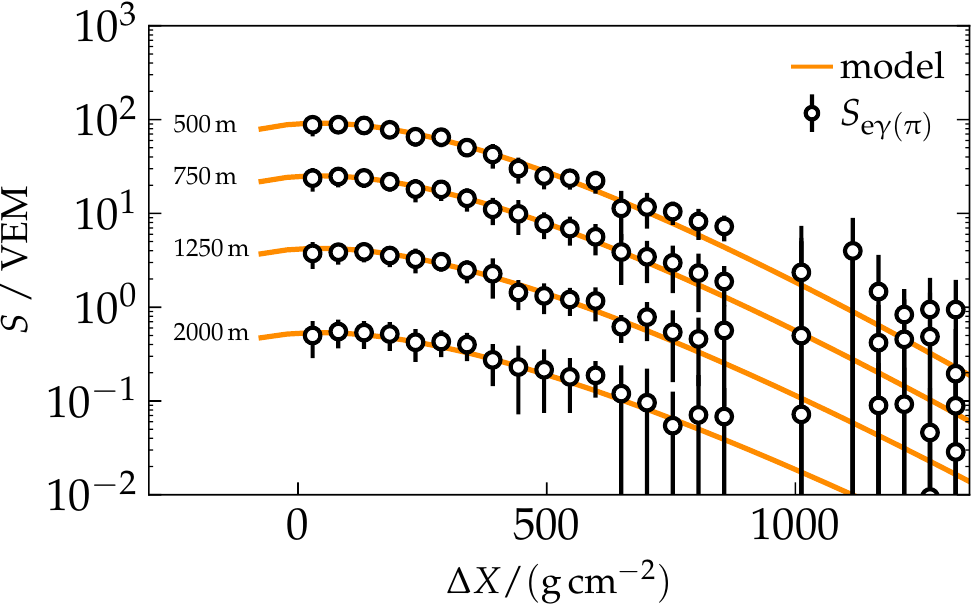}\hfill
    \includegraphics[width=\w\textwidth]{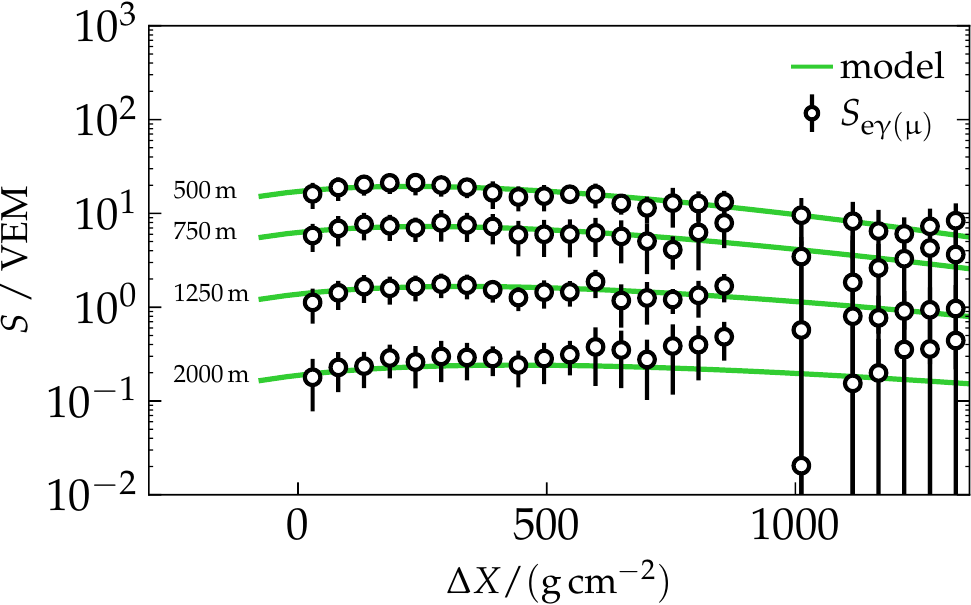}
    \caption{Average simulated signals, $S_\emcomp$, $S_\mucomp$, $S_\emhdcomp$, and $S_\emmucomp$ (points), of WCDs to air showers as a function of the distance $\DX$ to the shower maximum along with the corresponding model for the four particle components (lines).
    The signal is shown as a function of the distance $\DX$ to the shower maximum for four different radii 500\,m, 750\,m, 1250\,m, and 2000\,m.}
    \label{fig:s_ref}
\end{figure*}

\begin{figure*}
    \centering
    \def\w{0.46}
    \includegraphics[width=\w\textwidth]{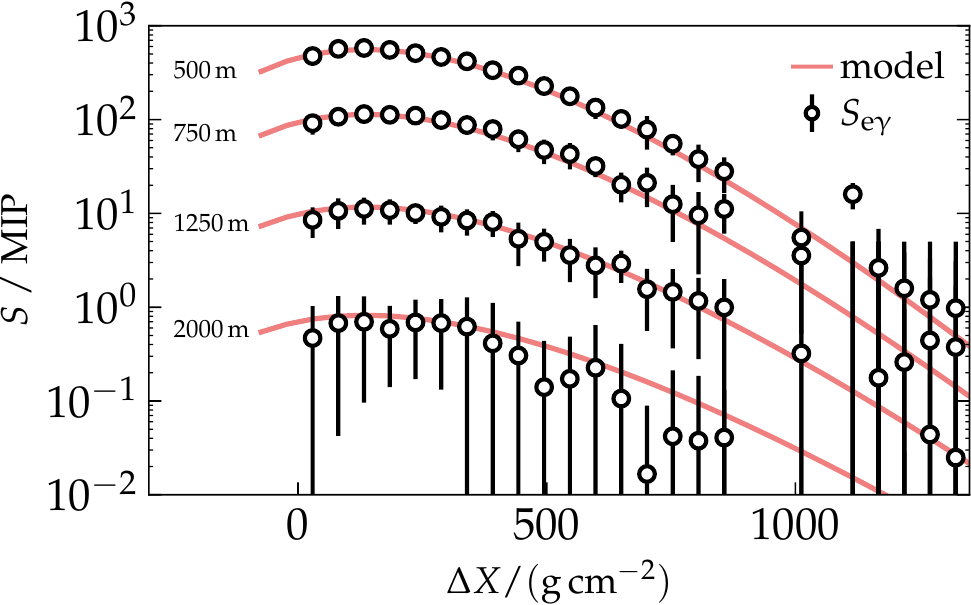}\hfill
    \includegraphics[width=\w\textwidth]{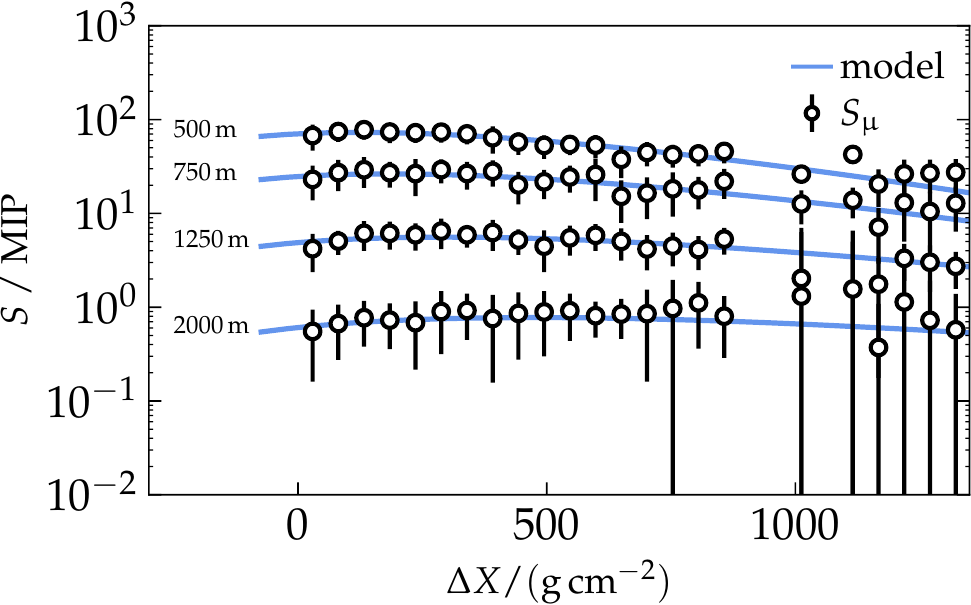}
    \\[3mm]
    \includegraphics[width=\w\textwidth]{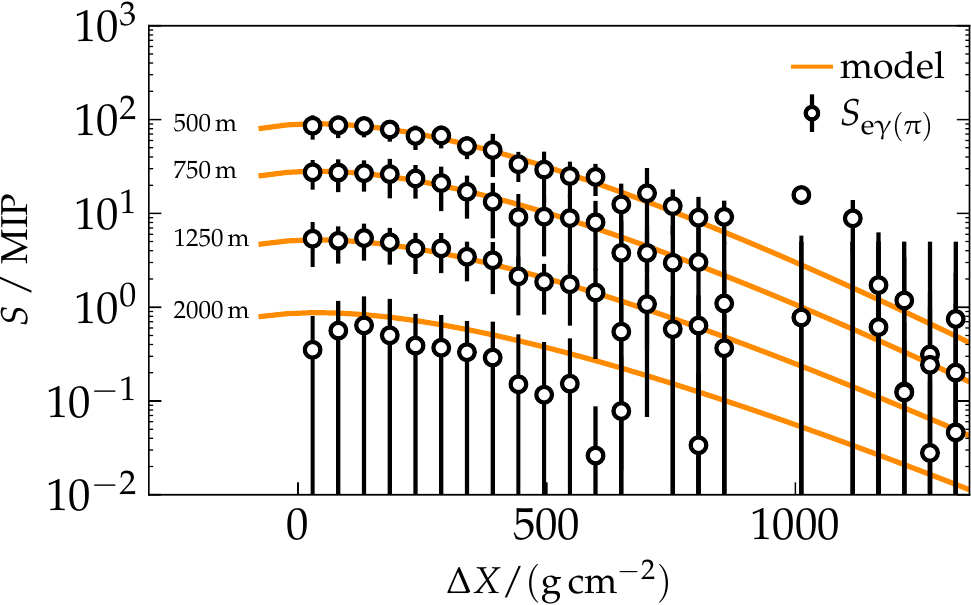}\hfill
    \includegraphics[width=\w\textwidth]{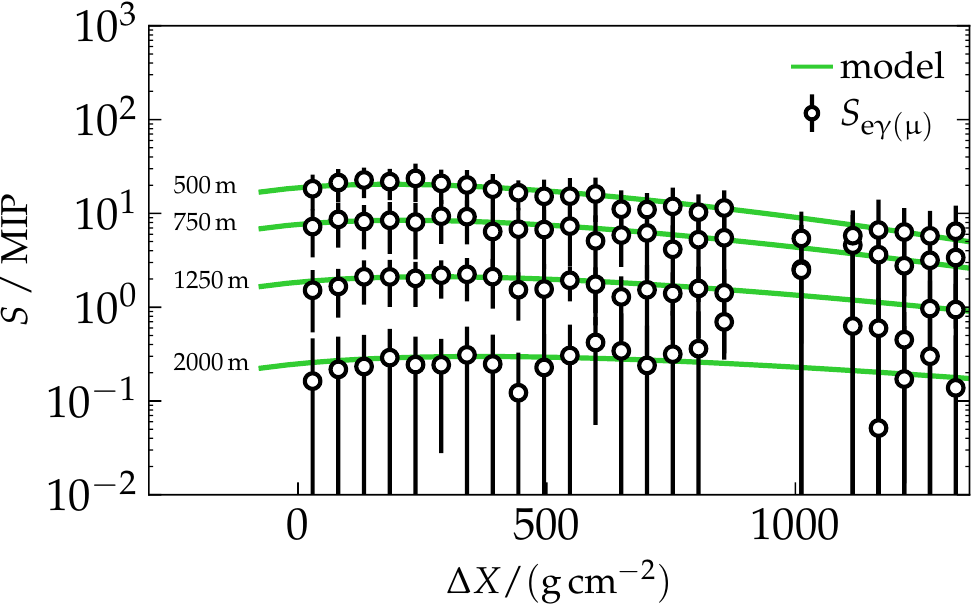}
    \caption{Average simulated signals, $S_\emcomp$, $S_\mucomp$, $S_\emhdcomp$, and $S_\emmucomp$ (points), of SSDs to air showers as a function of the distance $\DX$ to the shower maximum along with the corresponding model for the four particle components (lines). The signal is shown as a function of the distance to the shower maximum for four different radii 500\,m, 750\,m, 1250\,m, and 2000\,m.}
    \label{fig:s_ref_ssd}
\end{figure*}

\begin{figure*}
    \centering
    \def\w{0.45}
    \includegraphics[width=\w\textwidth]{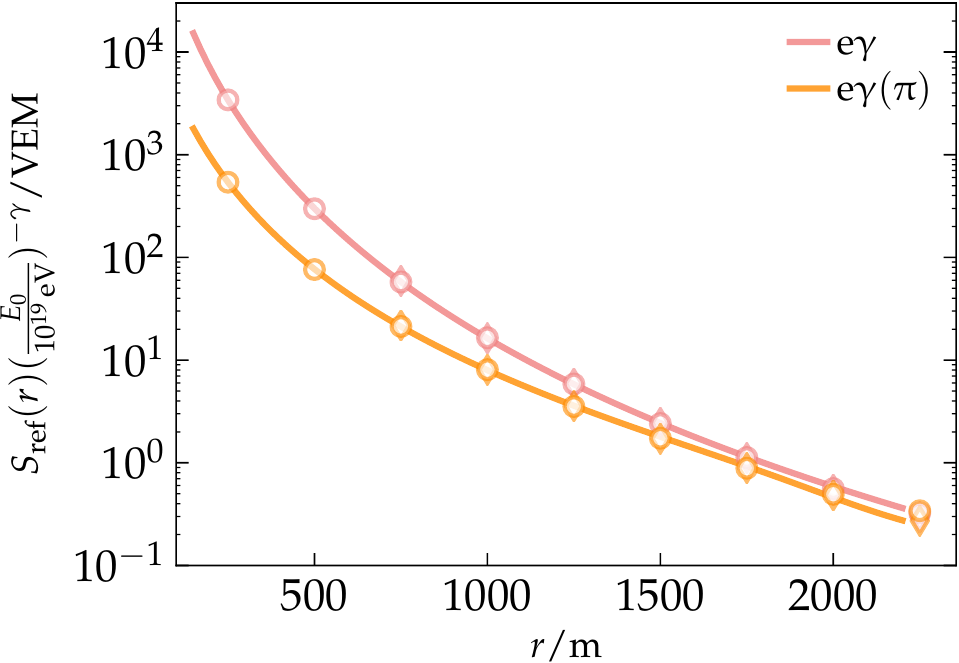}\hfill
    \includegraphics[width=\w\textwidth]{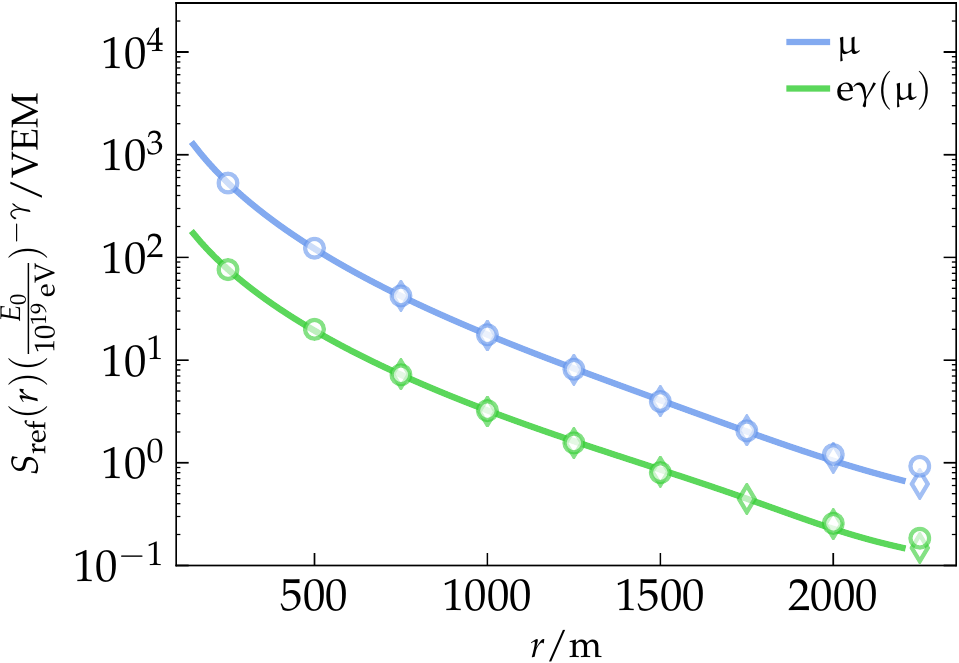}
    \\[3mm]
    \includegraphics[width=\w\textwidth]{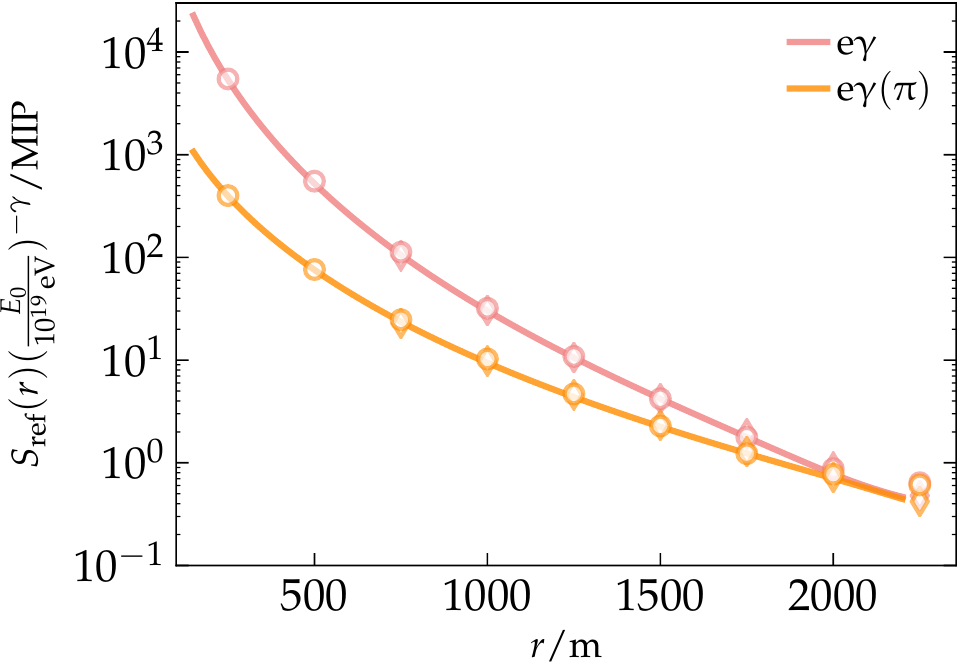}\hfill
    \includegraphics[width=\w\textwidth]{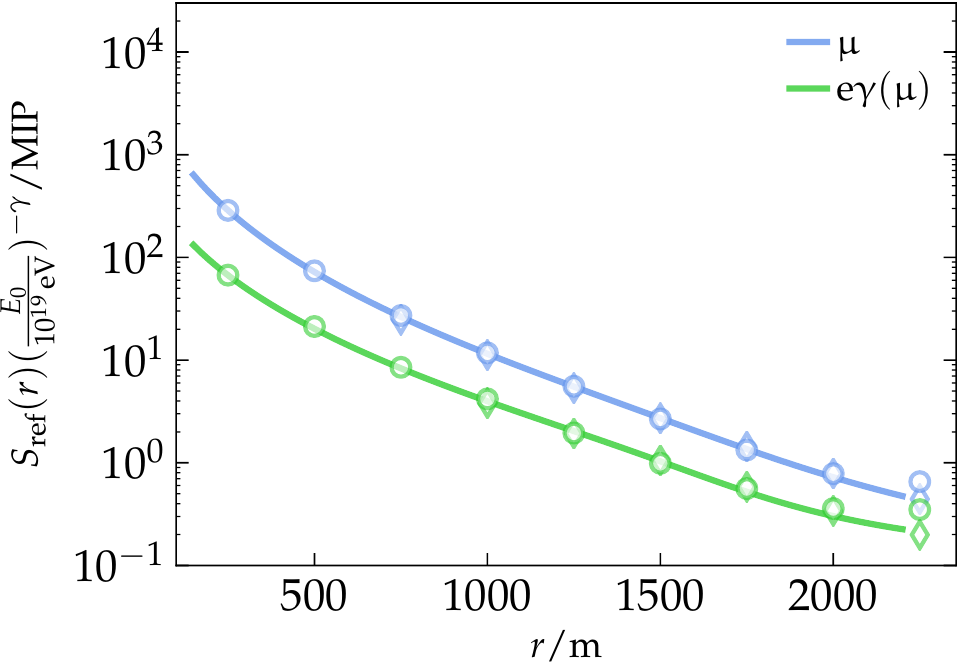}
    \caption{Lateral distribution $S_\text{ref}(r)$ of the average signal deposited in the WCDs (\emph{top row}) and SSDs (\emph{bottom row}) by the four particle components at the reference distance $\DX_\text{ref}=200\,\gcm$ from the shower maximum.
    The data are parametrized using \cref{eq:NKG} in terms of signal.
    Circular (diamond) markers show the lateral signal obtained from simulated proton showers with a primary energy of $10^{19}$\,eV ($10^{20}$\,eV) along with the corresponding model as a solid line.
    The energy dependence is removed by normalizing according to \cref{eq:rhoref}.}
    \label{fig:nkg}
\end{figure*}

\section{Parametrization of the expected signal in a surface-detector array}
\label{sec:param}

\begin{figure*}
    \centering
    \def\w{0.46}
    \includegraphics[width=\w\textwidth]{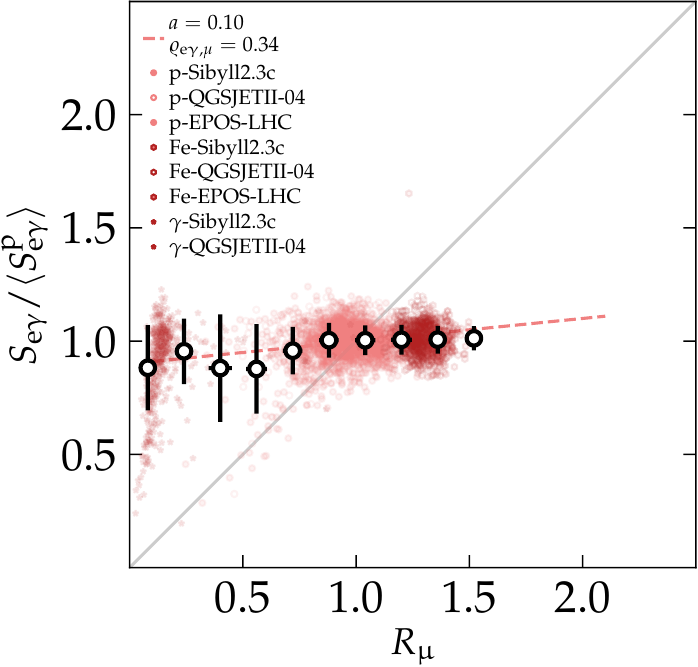}\hfill
    \includegraphics[width=\w\textwidth]{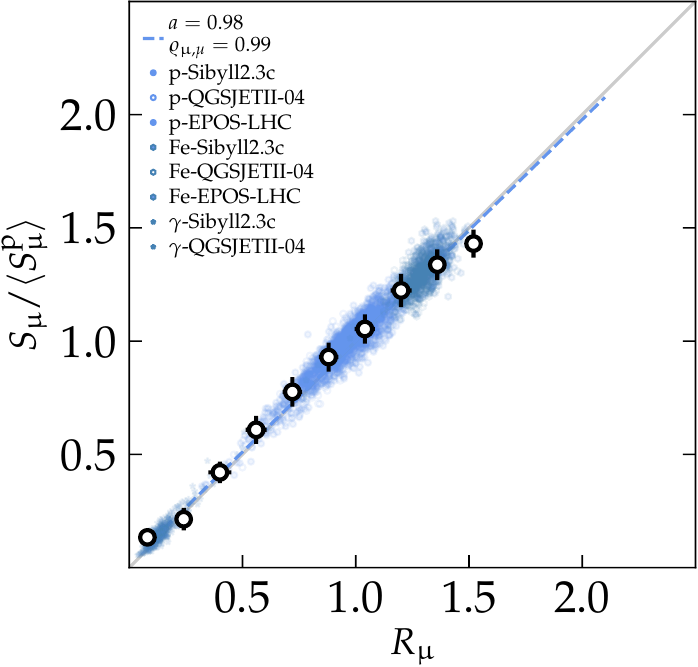}
    \\[3mm]
    \includegraphics[width=\w\textwidth]{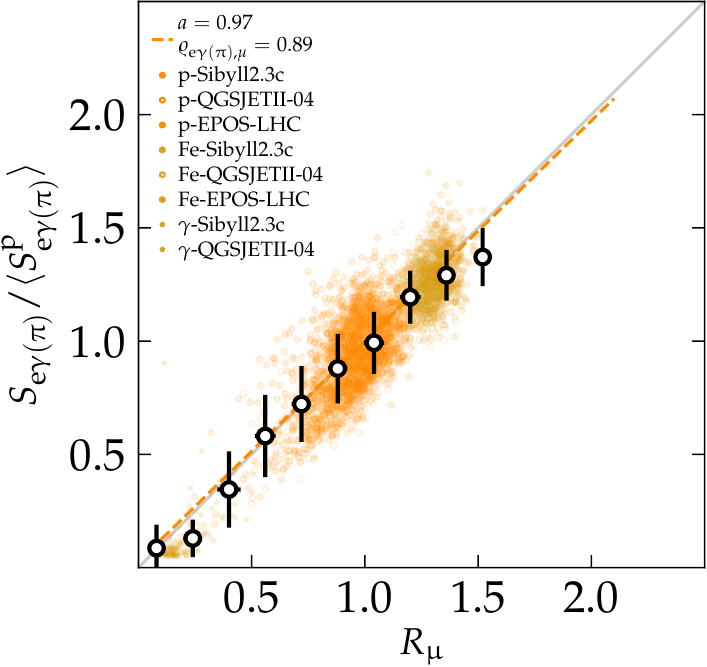}\hfill
    \includegraphics[width=\w\textwidth]{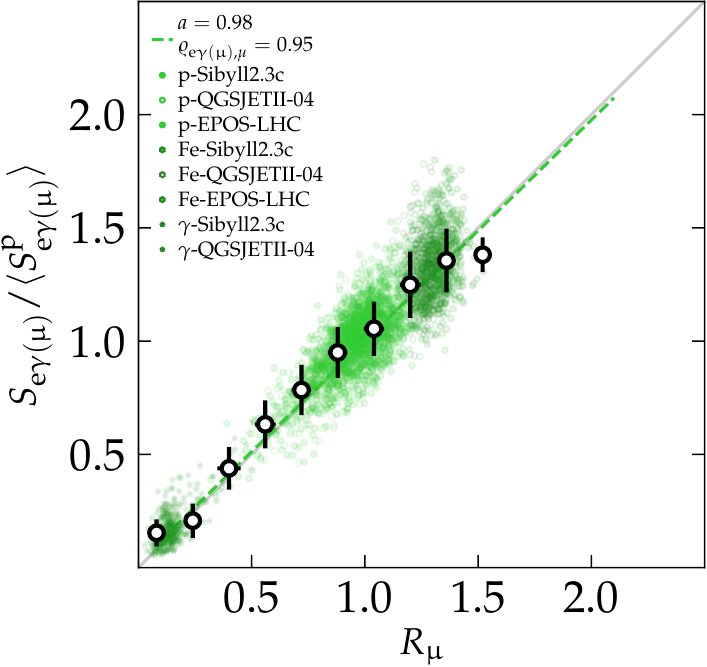}
    \caption{Normalized signal $S/\langle S^\text{p}\rangle$ of the particle components as a function of $\Rmu$.
    Data are obtained from simulated WCD signals at 750\,m, 1000\,m, and 1250\,m using different hadronic interaction models and primary particles.
    Showers were simulated with a primary energy $E_0=10^{19}$\,eV with zenith angles in a range from $0^\circ$ to $60^\circ$.
    For each component, \cref{eq:Ri} is expressed in terms of signal.
    The component scaling factor $a$ and the Pearson correlation coefficient $\varrho$ of the data are placed in the upper-left corner of each panel.}
    \label{fig:comp_corr}
\end{figure*}

Using simulations of the response of the upgraded SD array of the Pierre Auger Observatory, AugerPrime~\cite{PierreAuger:2016qzd}, produced in the \Offline software framework~\cite{Argiro:2007qg}, a model of the particle-component densities can be parameterized.
The model and parametrization described in this work is named \un2 and can be found as such within the \Offline software.
Parameterization is performed directly in terms of the signal $S$, so that no model of detector responses to individual particles is required.
For each type of detector and particle components, it is individually assumed that \cref{eq:masterequation} is valid directly in terms of the signal.
This assumption breaks down for very small signals at large distances from the shower maximum or from the shower axis, when the particle density is in the region of the detection threshold.

Simulated responses of the Auger water-Cherenkov detectors (WCDs) to proton showers with primary energies $10^{19}$\,eV are shown in \cref{fig:s_ref} for the four particle components individually, together with the fully parametrized model.
The signal is given in units of vertical equivalent muons (VEM)~\cite{PierreAuger:2007kus}.
Equivalent profiles for the model of the responses of scintillator surface detectors (SSDs) are shown in \cref{fig:s_ref_ssd}, with the signal given in units of the most probable signal charge of a vertical minimum-ionizing particle (MIP)~\cite{PierreAuger:2016qzd}.
The respective lateral profiles of the four components for both types of detectors are depicted in \cref{fig:nkg}.
The simulated air showers used to parameterize the model were produced using \textsc{Corsika}\,7~\cite{Heck:1998vt} employing the \textsc{Epos-LHC}~\cite{Pierog:2013ria} model of hadronic interactions.

The model depicted in \cref{fig:s_ref,fig:s_ref_ssd,fig:nkg} accurately describes the expected detector responses, except for very large $\DX$ and $r$, where the particle density becomes very low and Poissonian fluctuations as well as detector threshold effects start to dominate the behavior of the signals.
To better match the expected signal in the regions, where small signals are expected (at the order of magnitude of the detector threshold), a sigmoid-like function is multiplied to $f(r)$~\cite{stadelmaier:2022}; its effect is visible in \cref{fig:nkg}-bottom, especially for the $\emmucomp$ component at $r\simeq2000$\,m.
The model parameters for the four components of \textsc{Epos-LHC} protons are given in \cref{tab:params}.
Since the parametrization is performed in terms of signal rather than particle density, the interpretability of the individual parameters -- especially the shower age parameter $s$ -- is limited (similarly as the parametrizations found in Ref.~\cite{Linsley:1978xw}).

The model of the signal components shown in \cref{fig:s_ref} can be compared with data from simulated air-shower events using different hadronic interaction models and primary particles.
In this way, the validity of \cref{eq:Ri,eq:compcorrelation} can be tested.
The result is depicted in \cref{fig:comp_corr}.
The particle densities are estimated by the signal deposited in a WCD, using the signal expected from a proton shower as a reference.
The unambiguous dependence of the individual signal components on $\Rmu$, illustrated in \cref{fig:comp_corr}, shows that hadronic showers can be well described with a Universality-driven model by scaling the contribution of the individual components with $\Rmu$.
\cref{fig:comp_corr} implicitly acts as a validation of the model, as the data is centered around $1$ for the $\emcomp$ component, and around the identity for the other components.
The slight deviation of the slope (given in the legend as $a$) from 0 for the $\emcomp$ component is due to the fact that particles from photopion production and subsequent decay are not correctly accounted for as part of the electromagnetic shower.
These should be identified as a fifth shower component, $\upmu(\e\upgamma)$, in future work.

\begin{figure}
    \centering
    \def\w{0.9}
    \includegraphics[width=\w\columnwidth]{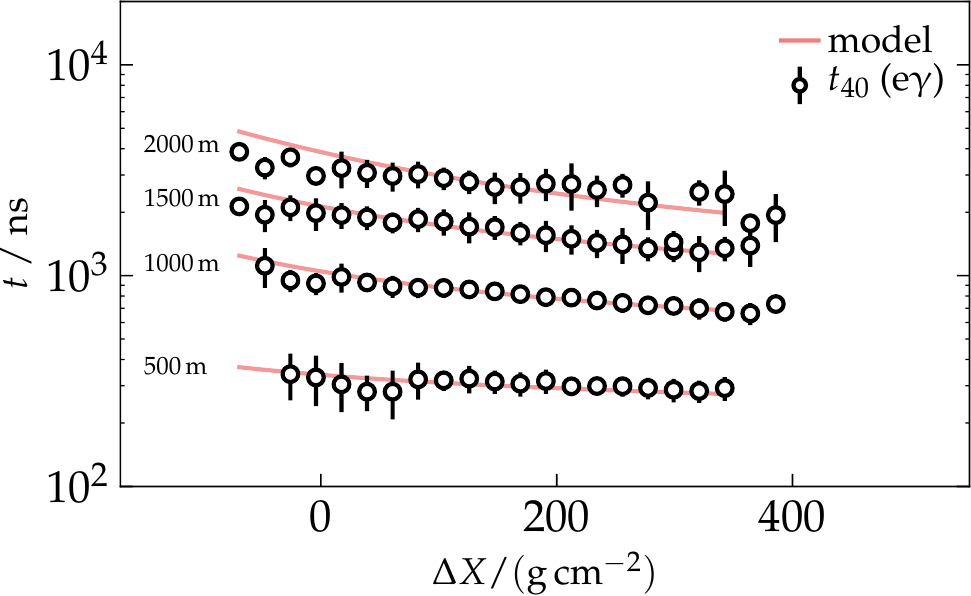}
    \\[3mm]
    \includegraphics[width=\w\columnwidth]{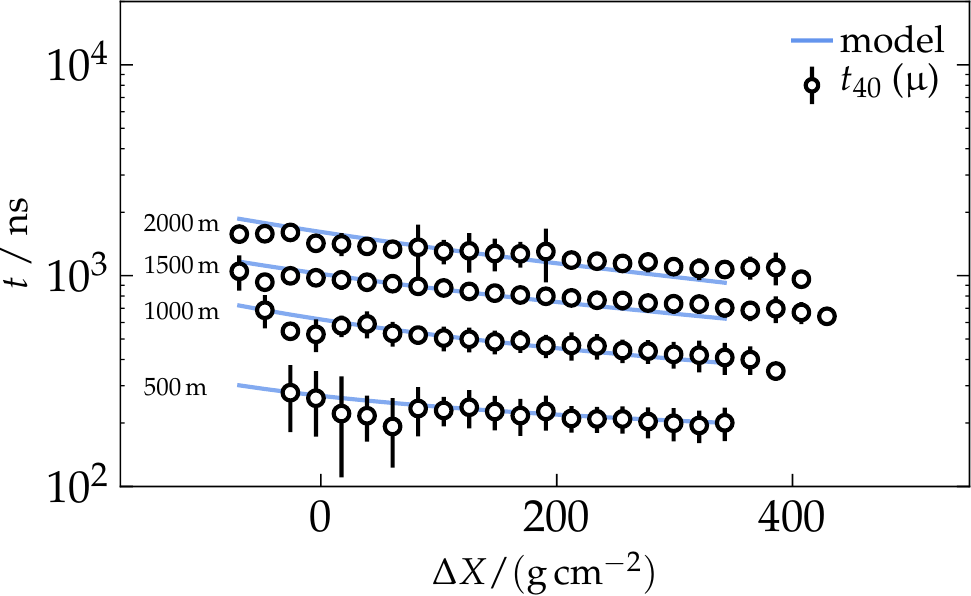}
    \\[3mm]
    \includegraphics[width=\w\columnwidth]{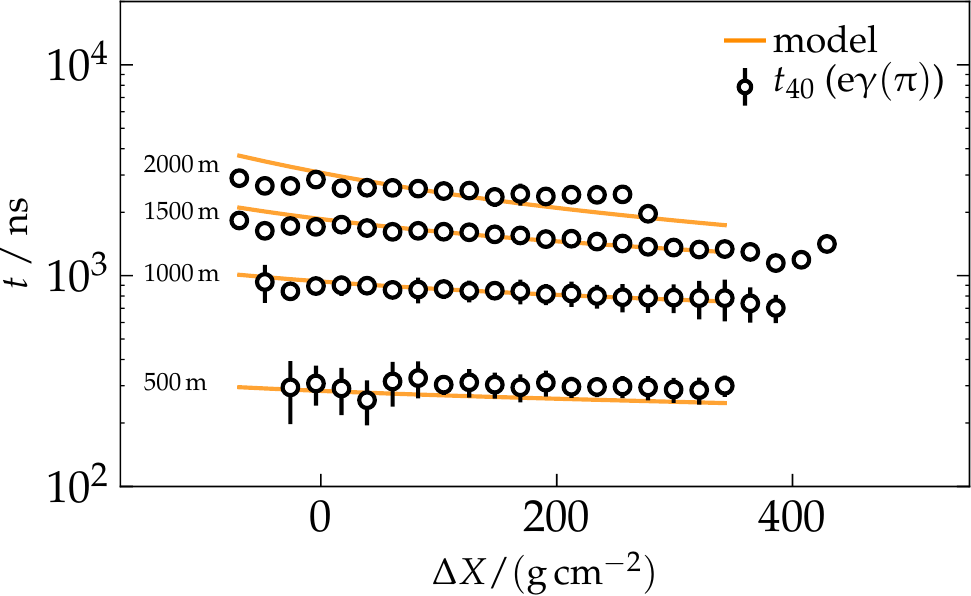}
    \\[3mm]
    \includegraphics[width=\w\columnwidth]{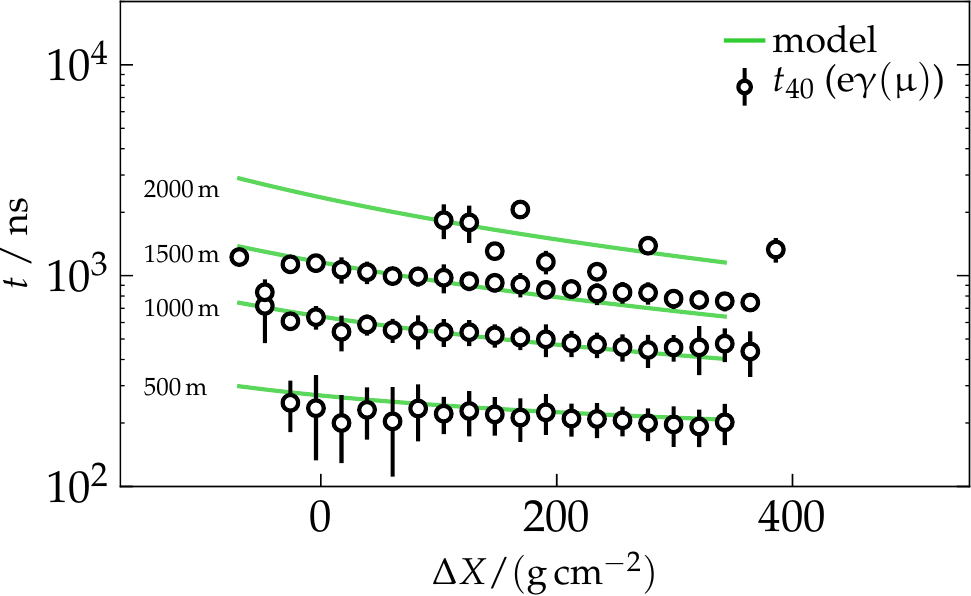}
    \caption{The average 40\% time quantile $t_{40}$ of simulated signals of individual detectors (points) as a function of the distance $\DX$ to the shower maximum, as well as the corresponding model for the four components (lines).
    The data are depicted for four sample radii.
    The data for stations at 2000\,m distance, where the model is not applicable anymore, is shown just for reference.}
    \label{fig:t40}
\end{figure}

The temporal distribution of particles is modelled using $t_{40}$ and $\sigma$ as obtained from simulated detector time traces, which were fit to a shifted log-normal distribution function.
\cref{eq:t40} can be parametrized as a function of $\DX$ for each of the four particle components, using a correction shift of the form $\DX \rightarrow \DX + \delta X$ and an additional term to take into account the real shape\footnote{The ideal shower front is a perfect expanding sphere, the real shape is, however, slightly parabolic.} of the shower front, $t_{40}\to t_{40} + t_\text{corr}$, where both $\delta X$ and $t_\text{corr}$ are parametrized as a function of $r$ and $\theta$~\cite{stadelmaier:2022}.
The correction terms $\delta X$ and $t_\text{corr}$ also account for the non-rectilinear propagation of the particles and the finite detector response time, respectively, which are otherwise not part of the model described in Section~\ref{subsec:time}.
Using a global best fit, it was evaluated that the model describes the simulated data best when using\footnote{Using $c < c_0$ captures possible kinematic delays and scattering.} $c = 0.95\,c_0$, with the speed of light $c_0$ in vacuum.
$t_{40}$ as a function of $\DX$ is shown in \cref{fig:t40} for four example radii.
The magnitude of the slope of $t_{40}$ as a function of $\DX$ shows the sensitivity of the \un2 model to the actual value of $\Xmax$ as seen in the time information from individual stations.
Only at large distances from the shower axis and/or the shower maximum (at $r\gtrsim1800$\,m and $\DX\gtrsim600\,\gcm$) the model breaks down due to the decreasing multiplicity of particles.
In these regions, the traces are not smooth anymore and $t_{40}$ shows no dependence on $\DX$.

In addition to a minor variation, the average of the shape parameter $\sigma$ is rather constant as a function of $\DX$, although with a large uncertainty.
The parametrization and the behaviour of the mean $\sigma$ as a function of the geometry is described in detail in Ref.~\cite{stadelmaier:2022}.

\begin{figure}
    \centering
    \def\w{0.9}
    \includegraphics[width=\w\columnwidth]{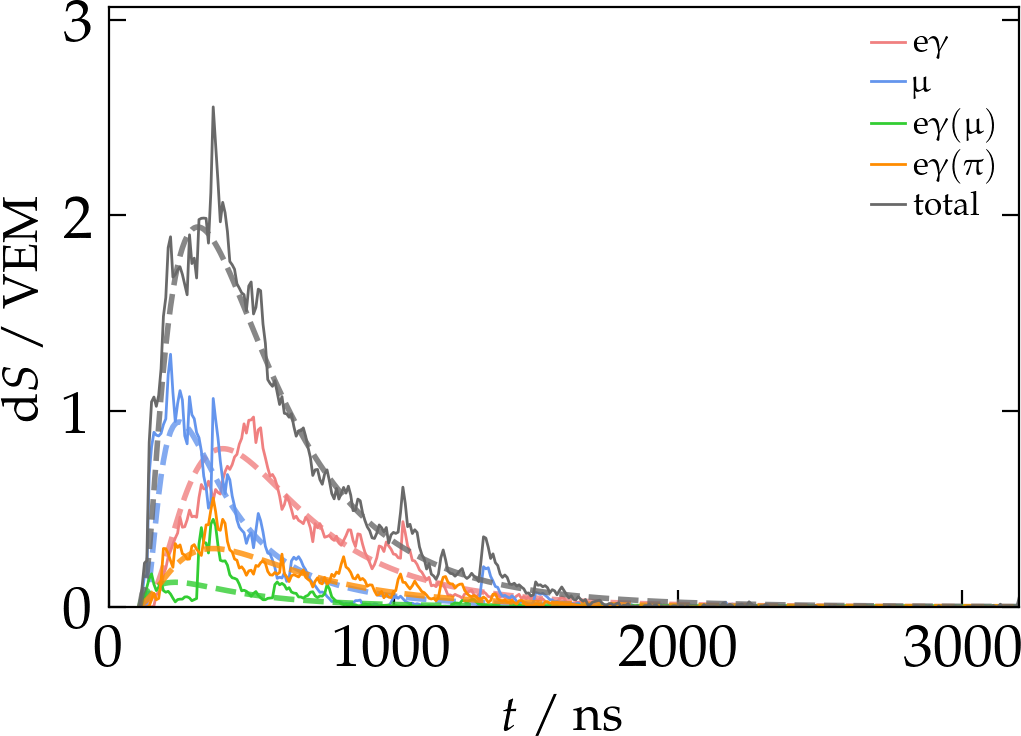}
    \\[3mm]
    \includegraphics[width=\w\columnwidth]{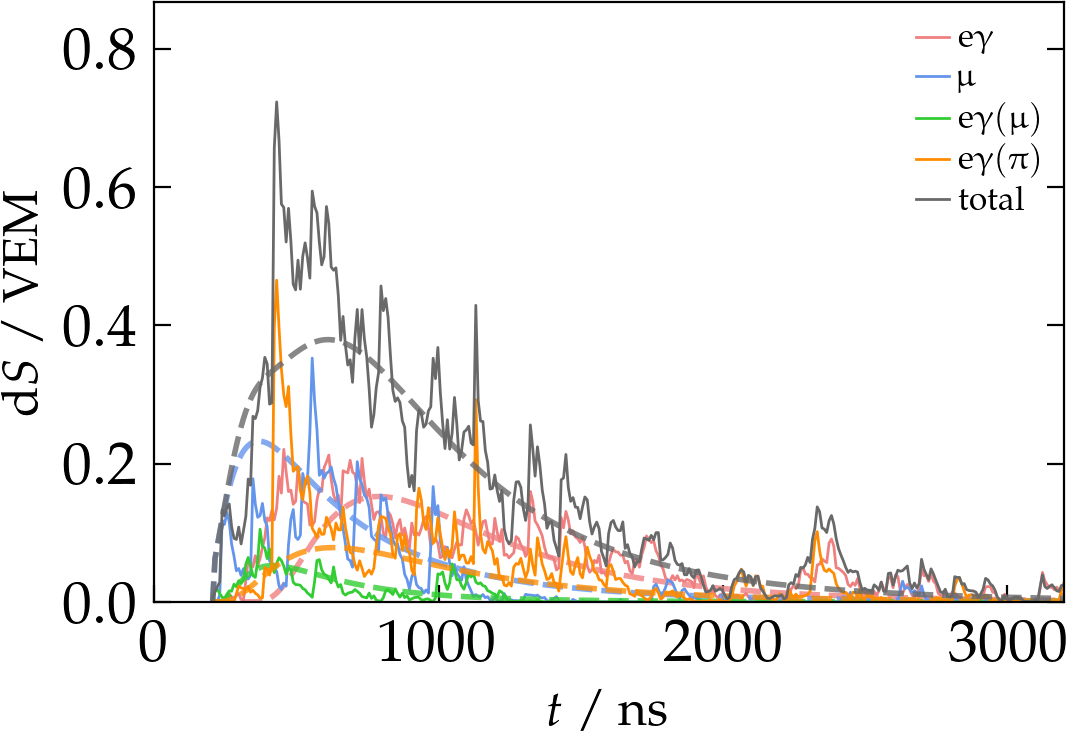}
    \\[3mm]
    \includegraphics[width=\w\columnwidth]{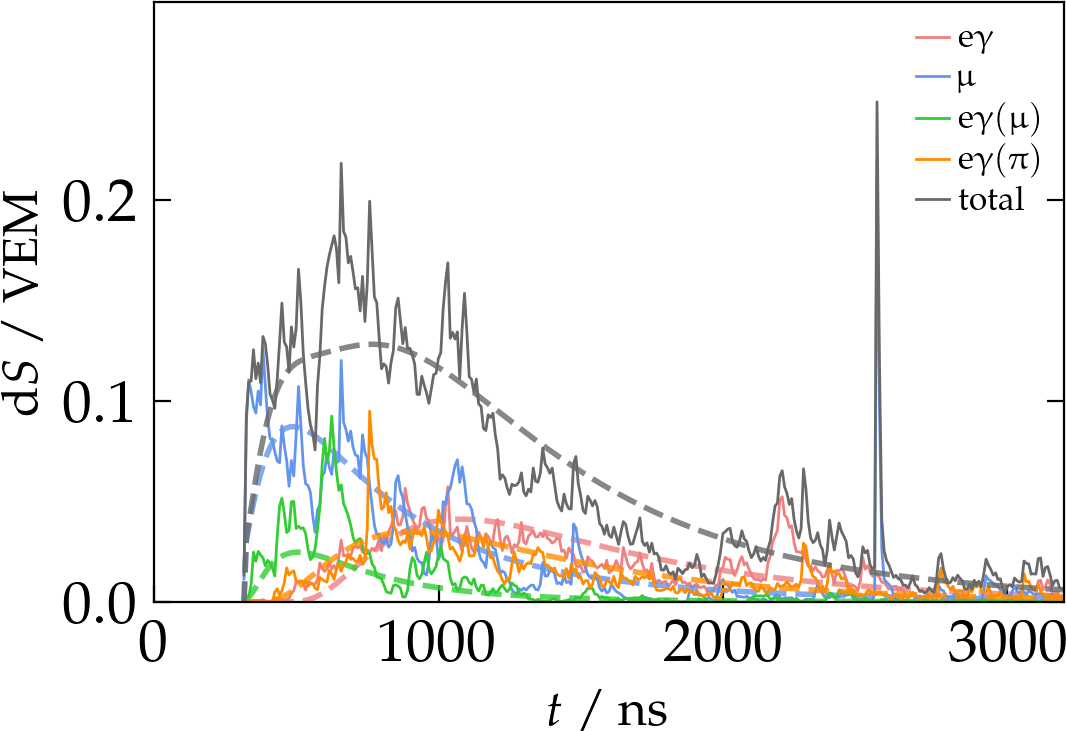}
    \caption{Visualisation of the model prediction for the time-dependent signal from a proton shower with primary energy $E_0=10^{19}$\,eV and zenith angle $\theta=22^\circ$ at three different distances $r=750$\,m (\emph{top}), 1000\,m (\emph{middle}), and 1250\,m (\emph{bottom}).
    The model prediction (dashed lines) for three example WCDs is shown along with simulated detector responses (thin solid lines) for the four components as well as for the total signal.
    $t=0$ marks the time of the plane front arriving at the position of the detector.}
    \label{fig:trace_example}
\end{figure}

The full prediction of the \un2 model for the response of three detectors to a simulated example event is shown in \cref{fig:trace_example}.
The model accurately describes the temporal distribution of the signal for the four components and its total, except for spikes of a single particles, as can be seen in the last panel of \cref{fig:trace_example}.

\section{Application of the Model}

\begin{figure}
    \centering
    \def\w{0.9}
    \includegraphics[width=\w\columnwidth]{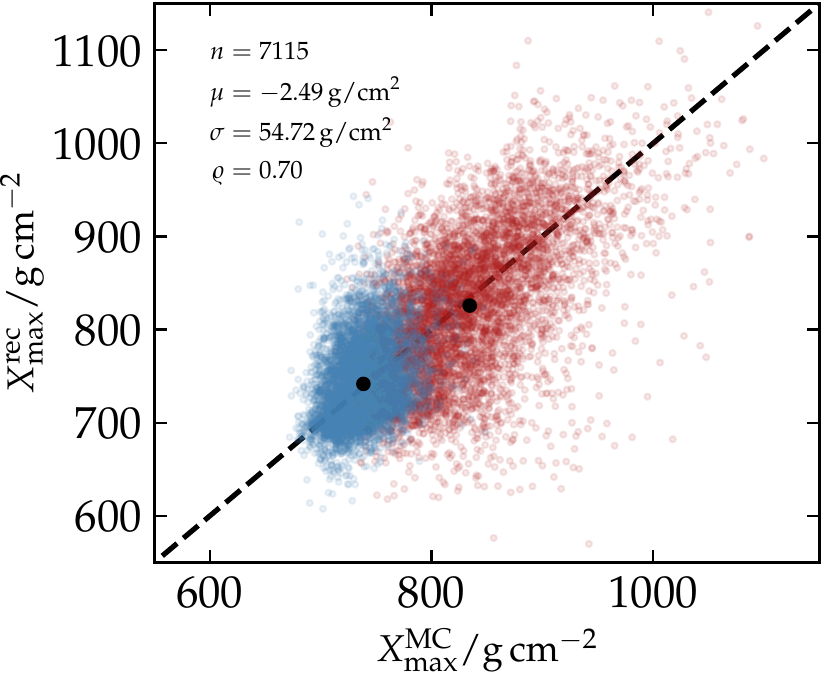}
    \\[3mm]
    \includegraphics[width=\w\columnwidth]{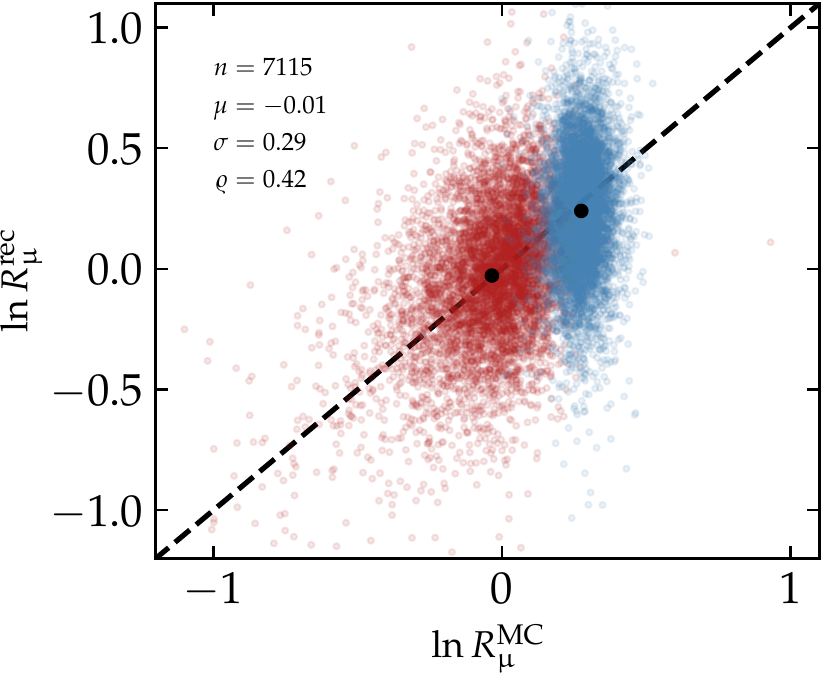}
    \\[3mm]
    \includegraphics[width=\w\columnwidth]{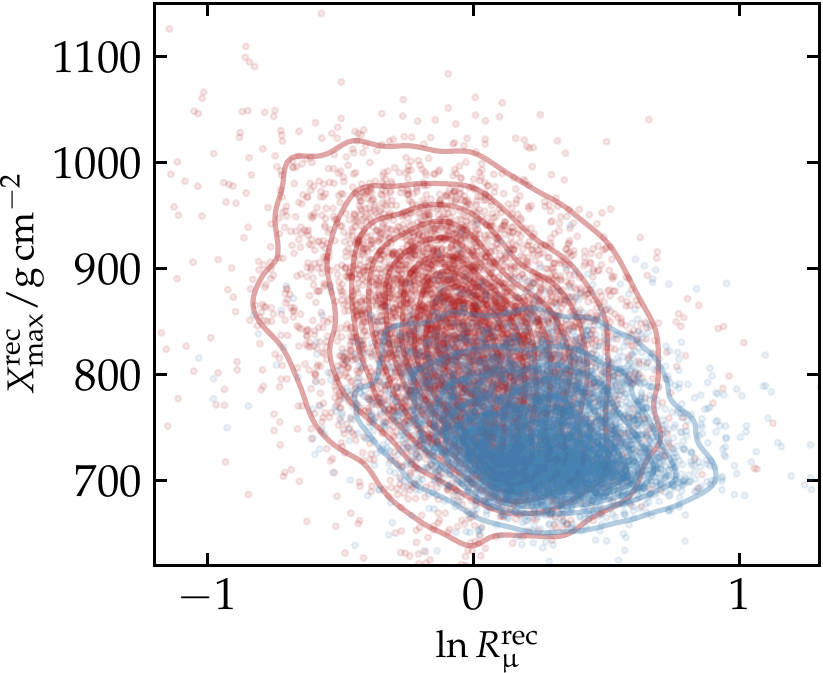}
    \caption{Simulated and reconstructed values of (\emph{top}) the depth of the shower maximum $X_\text{max}^\text{MC}$ and $X_\text{max}^\text{rec}$, and (\emph{middle}) scaling $\ln R_\upmu^\text{MC}$ and $\ln R_\upmu^\text{rec}$ for simulated air showers from primary iron particles (\emph{blue}) and primary proton (\emph{red}) with energies between $10^{18.5}$ and $10^{20}$\,eV with zenith angles in range from $0^\circ$ to $60^\circ$.
    The inset legend indicates the number of events $n$, accuracy $\mu$ and precision $\sigma$ of the reconstructed values, and the Pearson correlation $\rho$ of the bivariate data.
    Mean values for $\Xmax$ and $\ln\Rmu$ are indicated by black circular markers.
    The identity function is given as a dashed line in the top two plots.
    \emph{Bottom:} Two-dimensional distributions in reconstructed $\Xmax$ and $\ln\Rmu$ shown as individual data points alongside contour lines of estimated equal density.}
    \label{fig:rec}
\end{figure}

The \un2 model discussed in \cref{sec:param} can be used to reconstruct observables, most prominently $\Xmax$ and $\Rmu$, from SD data collected by the Pierre Auger Observatory.
Reconstruction of both $\Xmax$ and $\Rmu$ using a Universality-based model and only SD data has been a topic of research for many years~\cite{AVE201723,Ave201746}.
The main challenge of the estimation at event level of $\Rmu$ from SD data is the correlation of the number of muons $N_\upmu$ with the primary energy $E_0$ of the shower, the latter governing the total amount of particles produced.
An independent energy estimator, such as given by the Auger fluorescence detector or the SSDs, can help to disentangle this ambiguity.
Furthermore, as outlined in Section~\ref{sec:param}, only little information about $\Xmax$ is contained in SD data from a single detector station. 
At high energies, however, where many stations are triggered by a single event, $\Xmax$ can be reconstructed with reasonable accuracy and precision using the combined information of the station time traces (see \cref{fig:trace_example}).
The ability of the Universality-based reconstruction to estimate $\Xmax$ and $\Rmu$ is evaluated using simulations of the AugerPrime SD responses to the EAS produced by \textsc{Corsika}.
Reconstructed observables of showers are depicted in \cref{fig:rec}.
The results are obtained by minimizing an event-level likelihood function in terms of $\Xmax$ and $\Rmu$.
Stations that are expected to be located at a distance greater than $r=1800$, m or $\DX=1000\,\gcm$ from the shower axis or the maximum shower, respectively, are not considered in the fit for the reasons described in Section~\ref{sec:param}.
The reconstruction of these Monte-Carlo events was performed under the same conditions as later expected for measured data.
A global bias correction is performed for both the reconstructed $\Xmax$ and $\Rmu$ to accurately match the expectation values of showers simulated with the \textsc{Epos-LHC} model.
Since the mean values for $\Xmax$ (and $\Rmu$) for showers of different primary particles are subject to systematic uncertainties, arising from the differences in hadronic interaction models, it is foreseen to calibrate the reconstruction using the data from direct fluorescence detector measurements.
Note that the reconstruction of $\Xmax$ depicted here is based on only the WCD signal time traces, which are more smooth than the SSD signal time traces.
The reported performance is thus also expected from Auger SD data recorded before the upgrade of the SD.

At the energies considered, the average precision to reconstruct values of $\Xmax$, as depicted in \cref{fig:rec}, is approximately $55\,\gcm$ (slightly improving with increasing primary energy).
This is clearly no competition to direct shower profile observations provided by fluorescence detector measurements, which yield a precision of up to ${\approx}15\,\gcm$~\cite{PierreAuger:2014sui}.
This disadvantage is, however, countered by the largely increased statistics of SD measurements.
Furthermore, while a higher precision can be achieved using machine learning techniques~\cite{PierreAuger:2021fkf}, the \un2 model provides a \emph{classical} alternative to neural networks and its performance is well enough to distinguish a heavy and light component, given a mixed composition data set.
The same holds true for the relative number of muons, which can be on average reconstruction with a precision of about $25\%$.
This number could improve, for example, if novel methods to estimate the primary energy of UHECRs are successfully employed~\cite{PierreAuger:2023ohi}.
We consider the achieved precision reasonable, as it is smaller than the difference in expectation values of the respective observables for proton and iron primary particles (which in simulations is approximately $80\,\gcm$ in terms of $\Xmax$ and approximately $30\%$ in muons, depending on the underlying hadronic interaction model).
Moreover, because of shower-to-shower fluctuations, individual showers -- especially from light primary particles -- are expected to produce values in $\Xmax$ (and $\ln\Rmu$) that are by far larger (smaller) than the average of a given distribution of a mixed composition, as can be seen by the outlier data points shown in \cref{fig:rec} panels 1 and 2.
These can be easily recovered with the \un2 model and its given precision, if they are present in the reconstructed data.

\section{Discussion and summary}

In this article we presented a new model of SD responses to EAS from UHECRs.
The \un2 model is derived from the principle that the particle densities of air showers can be uniquely described based on the depth of the shower maximum, the number of muons in the shower, and the primary energy, independently of the nuclear mass of the primary cosmic ray.
Both the signal model and the time model are based on earlier work, but are thoroughly revisited and improved, especially in terms of the employed functional forms.
The time model is now based on considerations concerning the creation and propagation of the individual shower particles, and the signal model is formulated using common shower-profile functions, with a minimal set of free (fitted) parameters.
The model provides a complete and physically motivated picture of the expected particle densities in air-shower experiments.
Furthermore, the parametrization is performed for both the WCD and SSD detectors of the Pierre Auger Observatory using full detector simulations, while previous parametrizations were relying on tabulated detector responses.
Using the model realization for the SSDs, it is thus possible to describe also the data from the updated SD.
We show that the expected signals in simulated showers from different primary particles and different hadronic interaction models are well described.
Lastly, we demonstrate qualitatively that the model can be applied to air-shower data to estimate the number of muons and the depth of the shower maximum from the spatial and temporal distribution of particles at the ground with a reasonable precision for showers with a primary energy above 3\,EeV.
The application of the method will be performed in a forthcoming article.

\section{Acknowledgements}

The authors acknowledge fruitful discussions with colleagues of the Pierre Auger Collaboration, and helpful comments on the manuscript by Roger Clay.
We want to thank explicitly Johannes Hulsman, Federico Sanchez, Steffen Hahn, Jakub V\'icha, Eva Santos, and Alexey Yushkov for lively conversations in the context of this work. 
The work leading to this publication was partially supported by the PRIME programme of the German Academic Exchange Service (DAAD) and ErUM Pro (No.~05A20VK1) with funds from the German Federal Ministry of Education and Research (BMBF) and was partially supported by the Ministry of Education Youth and Sports of the Czech Republic and by the European Union under the grant FZU researchers, technical and administrative staff mobility, registration number CZ.02.2.69/0.0/0.0/18\_053/0016627 and grant project LM2023032.

\clearpage
\newpage

\bibliography{un2}{}

\appendix

\section{Relative rate of change of the longitudinal profile}
\label{app:rel}

The relative rate of change $\lambda_1$ of the longitudinal profile $g$ of a shower as a function of depth $t$ in radiation-length units, $t=X/X_0$, is given by
\begin{equation}
    \lambda_1 := \frac{1}{g(t)}\frac{\partial g(t)}{\partial t}.
    \label{eq:lambda1}
\end{equation}
Ref.~\cite{1956progress} introduces the approximation
\begin{equation}
    \lambda_1 \simeq \tfrac{1}{2}\left(s - 1 - 3\ln s \right),
\end{equation}
which, when inserted into \cref{eq:lambda1} and solved for $t$, results in the Greisen profile (interesting derivations of the Greisen profile are given in Refs.~\cite{Lipari:2008td,stadelmaier:2022}).
In case of the Gaisser-Hillas profile, the function $\lambda_1$ simply evaluates into
\begin{equation}
    \lambda_1 = -\frac{2(t - t_\text{max})}{3(t - t_1)}
\end{equation}
with $t_\text{max}=\Xmax/X_0$ and $t_1=X_1/X_0$. 
An exact calculation of $\lambda_1$ under Approximation B of the cascade equations is given in \cite{Lipari:2008td}.

\section{Universality of the electromagnetic component}
\label{app:emscale}

According to the Heitler-Matthews model of hadronic air showers \cite{Matthews:2005sd}, a shower induced by a primary particle with atomic mass $A$ and energy $E_0$ produces
\begin{equation}
N_\upmu^{(A)} = A^{1-\beta}\left(\frac{E_0}{\epsilon_\text{c}^\uppi}\right)^\beta
\end{equation}
muons, where $\epsilon_\text{c}^\uppi\simeq20$\,GeV is the critical energy, below which pions decay into muons, and $\beta\simeq0.95$ is the relative muon growth rate.
In this simplistic picture, at the decay stage of the shower development all the energy, available at the time in charged pions, is converted into muons.
Thus, in total the energy
\begin{equation}
E^{(A)}_{\upmu/\uppi} =
  \epsilon_\text{c}^\uppi \, N_\upmu^{(A)} =
  \epsilon_\text{c}^\uppi \, A^{1-\beta}\left(\frac{E_0}{\epsilon_\text{c}^\uppi}\right)^\beta
\end{equation}
is deposited into the muonic and hadronic component of the shower.
Suppose an extreme scenario, where none of the muons produced decays into the electromagnetic component.
We are left with $E_0=\epsilon_\text{c}^\emcomp\,N_\emcomp^{(A)} + E^{(A)}_{\upmu/\uppi}$ and thus
\begin{equation}
N_\emcomp^{(A)} =
  \left(E_0 - A^{1-\beta} E_{\upmu/\uppi}^{(1)}\right)\frac{1}{\epsilon_\text{c}^\emcomp}
\end{equation}
particles are being produced in the maximum of the $\emcomp$ component.
In this scenario, the relative change of $N_\emcomp^{(A)}$ as a function of $A$ is thus given as
\begin{equation}
\frac{1}{N_\emcomp^{(1)}}\frac{\partial N_\emcomp^{(A)}}{\partial A} = 
  A^{-\beta}
  (\beta-1)
  \frac{E_{\upmu/\uppi}^{(1)}}{ E_0 - E_{\upmu/\uppi}^{(1)}}
\approx
  -\frac{0.0125}{A^\beta},
\end{equation}
where the last expression is evaluated for $E_0\approx10^{19}$\,eV.
It is thus safe to assume that the number of particles in the $\emcomp$ component does not change significantly with the atomic mass $A$ of the primary particle for hadronic UHECRs.
The size of the \emph{pure} electromagnetic cascade is thus \emph{universal}.

If, however, a fraction $w$ of the muons and hadrons created in a shower decays into electromagnetic particles, one expects an proportional increase of the electromagnetic particles,
\begin{equation}
    \Delta N_{\emcomp(\upmu/\uppi)} = \frac{w\, E_{\upmu/\uppi}^{(A)}}{\epsilon^\emcomp_\text{c}}.
\end{equation}
In relative terms, for showers induced by proton primaries this makes up a a contribution of
\begin{equation}
  \frac{\Delta N_{\emcomp(\upmu/\uppi)}}
       {N_\emcomp^{(1)} + \Delta N_{\emcomp(\upmu/\uppi)}} =
  \frac{w\,E_{\upmu/\uppi}^{(A)}}
       {E_0 + (w-1)E_{\upmu/\uppi}^{(A)}} \approx
  \frac{w}{w + 1.7}.
\end{equation}
where, again, the last expression is evaluated for $E_0\approx10^{19}$\,eV.
A realistic estimate is $w\simeq 0.15$, so that in an ultra-high-energy proton shower ${\approx}10\%$ (and more so for heavier primary particles) of the electromagnetic particles are not part of the pure and universal electromagnetic component.
These particles follow different longitudinal and lateral distributions and therefore have to  be treated individually.
The number of these particles is expected to scale linearly with the number of muons produced in the shower.

\section{The distance to the shower maximum}

The distance $\DX$ of a SD station to the shower maximum is defined by the depth of the shower core, relative to $\Xmax$, when the plane-front shower is passing through the detector.
Using an isothermal approximation for the atmospheric density profile with a scale height $h_\text{s}$ and the total vertical depth of the ground $\Xvg$, the distance $\DX$ is given by
\begin{equation}
    \DX = \frac{X_\text{vg}}{\cos\theta}\,\exp{\left(-\frac{\hproj}{\hs}\right)} - \Xmax,
    \label{eq:un_dx_def}
\end{equation}
using the projected height of the shower core
\begin{equation}
    \hproj =r\cos\psi\,\sin\theta
    \label{eq:hproj}
\end{equation}
for a detector with the shower-plane coordinates $r$ and $\psi$.
To take into account the effects of a nonisothermal atmosphere, $\hs$ can be expanded in the first order as a function of $\hproj$,
\begin{equation}
    \hs = \hs^{(0)} + \hproj\,\hs^{(1)},
\end{equation}
with seasonally-dependent parameters $X_\text{vg}$, $\hs^{(0)}$, and $\hs^{(1)}$.

\section{Correction for azimuthal asymmetry}

We can model the azimuthal asymmetry for \cref{eq:NKG} as
\begin{equation}
u(\psi) =
  \exp\left[
    \zeta \;
    \left(\frac{\DX-\DX_1}{\lambda}\right)
    \left(\frac{r}{r_\text{ref}}\right)
    \cos\psi \,
    \sin\theta
  \right],
\label{eq:un_c_psi}
\end{equation}
using an arbitrary but fixed reference distance $r_\text{ref}=1000$\,m and a constant $\zeta \simeq \mathcal{O}(10^{-2})$ for each component.

\onecolumngrid

\section{Table of paramters}

\begin{table*}[hb!]
\caption{Parameters for the \texttt{un2} signal model based on shower simulations using proton primary particles and the \textsc{Epos-lhc} model.}
\label{tab:params}
\begin{ruledtabular}
\begin{tabular}{rllll}
        WCD parameters & $\emcomp$ & $\mucomp$ & $\emmucomp$ & $\emhdcomp$ \\
\hline
        $a$ & 0.0 & 1.0 & 1.0 & 1.0 \\
        $\gamma$ & 0.982 & 0.963 & 0.949 & 0.947 \\
        $N_{19}$ & $1.602{\times}10^{10}$ & $7.211{\times}10^8$ & $1.042{\times}10^8$ & $8.147{\times}10^8$ \\
        $r_\text{G}/\text{m}$ & 410 & 660 & 725 & 174 \\ 
        $s$ &  0.4 & 1.1 & 1.14 & 1.33 \\
        $\DX_1/(\gcm)$ & $-600$ & $-400$ & $-400$  & $-500$ \\
        $\DX_\text{max}/(\gcm)$ & 140 & $0.426(r/\text{m})$ & $154+0.141(r/\text{m})$ & 45 \\
        $\lambda/(\gcm)$ & $58+0.014(r/\text{m})$ & $280+0.201(r/\text{m})$ & $278+0.162(r/\text{m})$ & $98+0.011(r/\text{m})$ \\
        $\zeta$ & 0.03 & 0.01 & 0.03 & 0.1 \\
\hline\hline
\\[0.5em]
\hline\hline
        SSD parameters & $\emcomp$ & $\mucomp$ & $\emmucomp$ & $\emhdcomp$ \\
\hline
        $a$ & 0.0 & 1.0 & 1.0 & 1.1 \\
        $\gamma$ & 0.982 & 0.963 & 0.949 & 0.947 \\
        $N_{19}$ & $4.077{\times}10^{10}$ & $4.028{\times}10^8$ & $1.078{\times}10^8$ & $5.392{\times}10^8$ \\
        $r_\text{G}/\text{m}$ & 649 & 579 & 613 & 658 \\ 
        $s$ &  0.2 & 1.28 & 1.43 & 0.89 \\
        $\DX_1/(\gcm)$ & $-600$ & $-400$ & $-400$  & $-500$ \\
        $\DX_\text{max}/(\gcm)$ & 135 & $5+0.236(r/\text{m})$ & $121+0.122(r/\text{m})$ & 40 \\
        $\lambda/(\gcm)$ & $61+0.015(r/\text{m})$ & $250+0.187(r/\text{m})$ & $259+0.174(r/\text{m})$ & $111+0.019(r/\text{m})$ \\
        $\zeta$ & 0.03 & 0.01 & 0.03 & 0.1 \\
\end{tabular}
\end{ruledtabular}
\end{table*}

\end{document}